\documentclass[twocolumn,final,10pt]{elsarticle}

\usepackage[english]{babel} 

\usepackage[full]{textcomp}
\renewcommand{\copyright}{\textcopyright}

\usepackage{mathdesign}

\usepackage{multirow}
\usepackage{booktabs}
\usepackage{subcaption}
\usepackage[justification=centering]{caption}
\usepackage[dvipsnames]{xcolor}
\usepackage{pgfplots}
\usetikzlibrary{calc}

\usepackage{makecell, cellspace, caption}

\usepackage{amssymb} 
\usepackage{amsfonts}
\usepackage{amsmath}
\usepackage{amsthm}
\usepackage{amsmath,amsfonts,amsthm,bm} 
\usepackage{hyperref} 
\hypersetup{
        colorlinks
}
\usepackage{algorithm} 
\usepackage{algpseudocode} 

\usepackage[switch,displaymath]{lineno}

\journal{Medical Image Analysis}

\title{Segmentation with mixed supervision: 
Confidence maximization helps knowledge distillation}

\newcommand{\mr}[1]{\mathrm{#1}}

\newcommand{\mypar}[1]{\vspace{.5em}\noindent\textbf{#1}~}
\newcommand{\XX}{\mathbf{X}}
\newcommand{\YY}{\mathbf{Y}}

\newcommand{\yy}{\mathbf{y}}
\newcommand{\pp}{\mathbf{p}}
\newcommand{\qq}{\mathbf{q}}
\newcommand{\PP}{\mathbf{P}}

\newcommand{\ee}{\bm \epsilon}
\newcommand{\real}{\mathbb{R}}

\algnewcommand\algorithmicinput{\textbf{Input:}}
\algnewcommand\Input{\item[\algorithmicinput]}
\algnewcommand\algorithmicoutput{\textbf{Output:}}
\algnewcommand\Output{\item[\algorithmicoutput]}

\author[1]{Bingyuan Liu\corref{cor1}}
\cortext[cor1]{Corresponding author: \url{bingyuan.Liu@etsmtl.ca}}

\author[1]{Christian Desrosiers}

\author[1,2]{Ismail Ben Ayed}

\author[1,2]{Jose Dolz\corref{cor2}}
\cortext[cor2]{Corresponding author: \url{jose.dolz@etsmtl.ca}}

\address[1]{ÉTS Montréal}
\address[2]{Centre de recherche du Centre hospitalier de l’Université de Montréal (CRCHUM)}

\newcommand{\extension}[1]{{\color{black}#1}}
\newcommand{\rev}[1]{{\color{black}#1}}
\newcommand{\revA}[1]{{\color{black}#1}}
\newcommand{\revB}[1]{{\color{black}#1}}

\begin{document}
\begin{frontmatter}
\begin{abstract}
\extension{Despite achieving promising results in a breadth of medical image segmentation tasks}, deep neural networks (DNNs) require large training datasets with pixel-wise annotations. \extension{Obtaining these curated datasets is a cumbersome process which limits the applicability of DNNs in scenarios where annotated images are scarce. Mixed supervision is an appealing alternative for mitigating this obstacle. In this setting, only a small fraction of the data contains complete pixel-wise annotations and other images have a weaker form of supervision}, e.g., only a handful of pixels are labeled. In this work, we propose a dual-branch architecture, where the upper branch (teacher) receives strong annotations, while the bottom one (student) is driven by limited supervision and guided by the upper branch. Combined with a standard cross-entropy loss over the labeled pixels, our novel formulation integrates two important terms: (i) a Shannon entropy loss defined over the less-supervised images, which encourages confident student predictions in the bottom branch; and (ii) a Kullback-Leibler  (KL)  divergence term, which transfers the knowledge (i.e., predictions) of the strongly supervised branch to the less-supervised branch and guides the entropy (student-confidence) term to avoid trivial solutions. We show that the synergy between the entropy and KL divergence yields substantial improvements in performance. We also discuss an interesting link between Shannon-entropy minimization and standard pseudo-mask generation, and argue that the former should be preferred over the latter for leveraging information from unlabeled  pixels. We evaluate the effectiveness of the proposed formulation through a series of quantitative and qualitative experiments using \extension{two publicly available datasets}. 
Results demonstrate that our method significantly outperforms other strategies for semantic segmentation within a mixed-supervision framework, \extension{as well as recent semi-supervised approaches}. Moreover, in line with recent observations in classification, we show that the branch trained with reduced supervision and guided by the top branch largely outperforms the latter. Our code is publicly available: \url{https://github.com/by-liu/ConfKD}.
\end{abstract}


\end{frontmatter}

\setlength{\parskip}{3pt}

\section{Introduction}

The advent of deep learning has led to the emergence of high-performance models which \extension{currently dominate the medical image segmentation literature} \citep{litjens2017survey,dolz20183d,ronneberger2015u}. \extension{The availability of large training datasets with high-quality segmentation ground-truth has been a key factor for these advances.} Nevertheless, obtaining such annotations is a cumbersome process prone to observer variability, \extension{which is further magnified when volumetric data is involved}. To alleviate the need for large labeled datasets, weakly supervised learning has recently emerged as an appealing alternative. In this scenario, one has access to a large amount of weakly labeled data that can come in the form of bounding boxes \citep{kervadec2020bounding,rajchl2016deepcut}, scribbles \citep{lin2016scribblesup}, image tags \citep{lee2019ficklenet} or anatomical priors \citep{kervadec2019constrained,peng2020discretely}. However, even though numerous attempts have been made to train segmentation models from weak supervision, most of them still fall behind their supervised counterparts, limiting their applicability in real-world settings. 

Another promising learning scenario is mixed supervision, where only a small fraction of data is densely annotated and a larger dataset contains less-supervised images. In this setting, which helps keeping the annotation budget under control, strongly-labeled data -- where all pixels are annotated -- can be combined with images presenting weaker forms of supervision. 
Prior literature \citep{lee2019ficklenet,rajchl2016deepcut} has focused mainly on leveraging weak annotations to generate accurate initial pixel-wise annotations, or \textit{pseudo-masks}, which are then combined with strong types of supervision to augment the training dataset. The resulting dataset is employed to train a segmentation network, mimicking fully supervised training. Nevertheless, we argue that treating both equally in a single branch may result in limited improvements, as the less-supervised data is underused. Other approaches resort to multi-task learning \citep{mlynarski2019deep,shah2018ms,wang2019mixed}, where the mainstream task (i.e., segmentation) is assisted by auxiliary objectives that are typically integrated in the form of localization or classification losses.
While multi-task learning might enhance the common representation for both tasks in the feature space, this strategy has some drawbacks. First, the learning of relevant features is driven by commonalities between the multiple tasks, which may generate suboptimal representations for the mainstream task. Secondly, having distinct task-objectives ignores the direct interaction between the multi-stream outputs, for example, explicitly enforcing consistency between the predictions of multiple branches. As we show in our experiments, considering such interaction significantly improves the results.

Motivated by these observations, we propose a novel formulation for learning with mixed supervision in medical image segmentation. Particularly, our dual-branch network imposes a separate processing of the strong and weak annotations, which prevents direct interference of different supervision cues. This is supported by the recent findings in \citep{luo2020semi}, who demonstrated empirically that bundling different forms of supervision together to train a segmentation network is problematic, and argued that joint treatment of different supervision under-exploits less supervised samples, introducing limited improvement. In particular, authors pointed out two key issues related to equal treatment of different levels of supervision: \textit{sample imbalance} (commonly much more less-supervised samples than fully supervised ones) and \textit{supervision inconsistency} (less-supervised samples provide lower quality annotations). The former introduces a high risk of overfitting towards less supervised data, whereas the later induces inconsistencies in the supervisory signals. Authors validated these hypothesis in their experiments, and showed that by decoupling branches receiving different types of labels the supervision inconsistency and biases from class imbalance can be eliminated.

As the uncertainty of predictions for unlabeled pixels can be high, the proposed model includes a loss term based Shannon entropy that enforces high-confidence predictions over the whole image. Moreover, in contrast to prior works in mixed-supervision \citep{mlynarski2019deep,shah2018ms,wang2019mixed}, which have overlooked the co-operation between multiple branches by considering independent multi-task objectives, we introduce a Kullback-Leibler (KL) divergence term. The benefits of the latter are two-fold. First, it transfers the predictions generated by the strongly supervised branch (teacher) to the less-supervised branch (student). Second, it guides the entropy (student-confidence) term to avoid trivial solutions.
Interestingly, we show that the synergy between the entropy and KL term yields substantial improvements in performances. Furthermore, we discuss an interesting link between Shannon-entropy minimization  
and pseudo-mask generation, and argue that the former should be preferred over the latter for leveraging information from unlabeled pixels. We report comprehensive experiments and comparisons with other strategies for learning with mixed supervision, which show the effectiveness of our novel formulation. An interesting finding is that the branch receiving weaker supervision considerably outperforms the strongly supervised branch. This phenomenon, where \textbf{the student surpasses the teacher's performance}, is in line with recent observations in the context of image classification \citep{furlanello2018born,yim2017gift}. 

\extension{A preliminary conference version of this work has appeared at IPMI'21 \citep{dolz2021teach}. Nevertheless, this journal version provides a substantial extension. First, we further discuss the current literature in semi-supervised segmentation, which is closely related to the proposed methodology. Furthermore, we have performed several additional experiments to demonstrate the robustness and usability of our approach. In particular, new main experiments include: 1) benchmark against well-known and recent semi-supervised segmentation methods, 2) evaluation of our model in the publicly available Left Atrium (LA) segmentation challenge, 3) assessing the impact of several components in the methodology and 4) studying the impact of alternative divergence functionals as consistency terms in our formulation. In addition to the theoretical insights regarding the preference of directly minimizing the entropy of the predictions over using pseudo-labels given in the conference version, we provide empirical evidence that employing pseudo-labels has indeed a strong pushing effect on uncertain predictions at the beginning of the training. }

\section{Related work}

\mypar{Mixed-supervised segmentation} An appealing alternative to training CNNs with large labeled datasets is to combine a reduced number of fully-labeled images with a larger set of images with reduced annotations. These annotations can come in the form of bounding boxes, scribbles or image tags, for example\footnote{Note that this type of supervision differs from semi-supervised methods, which leverage a small set of labeled images and a much larger set of unlabeled images.}. A large body of the literature in this learning paradigm addresses the problem from a multi-task objective perspective \citep{hong2015decoupled,bhalgat2018annotation,shah2018ms,mlynarski2019deep,wang2019mixed}, which might hinder their capabilities to fully leverage joint information for the mainstream objective. 
Furthermore, these methods typically require carefully-designed task-specific architectures, which also integrate task-dependent auxiliary losses, limiting the applicability to a wider range of annotations. For example, the architecture designed in \citep{shah2018ms} requires, among others, landmark annotations, which might be difficult to obtain in many applications. More recently, Luo et \textit{al.} \citep{luo2020semi} promoted the use of a dual-branch architecture to deal separately with strongly and weakly labeled data. Particularly, while the strongly supervised branch is governed by available fully annotated masks, the weakly supervised branch receives supervision from a proxy ground-truth generator, which requires some extra information, such as class labels. While we advocate the use of independent branches to process naturally different kinds of supervision, we believe that this alone is insufficient, and may lead to suboptimal results. Thus, our work differs from \citep{luo2020semi} in several aspects. First, we make a better use of the labeled images by enforcing consistent segmentations between the strongly and weakly supervised branches on these images. 
Furthermore, we enforce confident predictions at the weakly supervised branch by minimizing the Shannon entropy of the softmax predictions. 

\extension{
\mypar{Semi-supervised segmentation in medical images} Semi-supervised learning is closely related to the proposed methodology. In this scenario, a small number of labeled images are leveraged with a much larger set of unlabeled images. In recent years, a breadth of semi-supervised approaches have been proposed for medical image segmentation, including techniques based on adversarial learning~\citep{zhang2017deep}, self-training~\citep{bai2017semi}, manifold learning~\citep{baur2017semi},
 co-training~\citep{peng2020deep,zhou2019semi}, temporal ensembling~\citep{perone2018deep}, data augmentation~\citep{chaitanya2019semi}, consistency regularization~\citep{bortsova2019semi} and mutual information maximization~\citep{Peng2021boost}. The common element of these approaches is adding an unsupervised loss computed on unlabeled images, which regularizes the learning. In contrast, our model exploits images that can be fully or partly annotated, processing each type in a separate branch of the proposed network.
}

\mypar{Distilling knowledge in semantic segmentation} Transferring knowledge from one model to another has recently gained attention in segmentation tasks. For example, the teacher-student strategy has been employed in model compression \citep{bar2019robustness}, to distil knowledge from multi-modal to mono-modal segmentation networks \citep{hu2020knowledge}, or in domain adaptation \citep{xu2019self}. 
Semi-supervised segmentation has also benefited from teacher-student architectures \citep{cui2019semi,sedai2019uncertainty}. In these approaches, however, the segmentation loss evaluating the consistency between the teacher and student models is computed on the unannotated data. A common practice, for example, is to add additive Gaussian noise to the unlabeled images, and enforce similar predictions for the original and noised images. This contrasts with our method, which enforces consistency only on the strongly labeled data, thereby requiring less additional images to close the gap with full supervision.

\section{Methodology}


\extension{We first define} the set of training images as $\mathcal{D} = \{(\XX_n, \YY_n)\}_n$, where $\XX_i \in \real^{\Omega_i}$ represents the \textit{i}$^{th}$ image and $\YY_i \in \{ 0,1 \}^{\Omega_i \times C}$ its corresponding ground-truth segmentation mask. $\Omega_i$ denotes the spatial image domain and $C$ the number of segmentation classes (or regions). We assume the dataset has two subsets: $\mathcal{D}_s=\{(\XX_1,\YY_1),...,(\XX_m,\YY_m)\}$, which contains complete pixel-level annotations of the associated $C$ categories, and $\mathcal{D}_w=\{(\XX_{m+1},\YY_{m+1}),...,(\XX_n,\YY_n)\}$, whose labels can take the form of semi- or weakly-supervised annotations \rev{(e.g., scribbles, points, bounding boxes or image-tags)}. Furthermore, for each image $\XX_i$ in $\mathcal{D} = \mathcal{D}_s \cup \mathcal{D}_w$, $\PP_i \in [ 0,1 ]^{\Omega_i \times C}$ denotes the softmax probability outputs of the network, i.e., the matrix containing a simplex column vector $\pp_i^l = \left ( p_i^{l,1}, \dots, p_i^{l,C} \right )^{T} \in [0, 1]^C$ for each pixel $l \in \Omega_i$. Note that we omit the parameters of the network here to simplify notation. 

\begin{figure*}[t]
    \centering
    \includegraphics[width=0.8\textwidth]{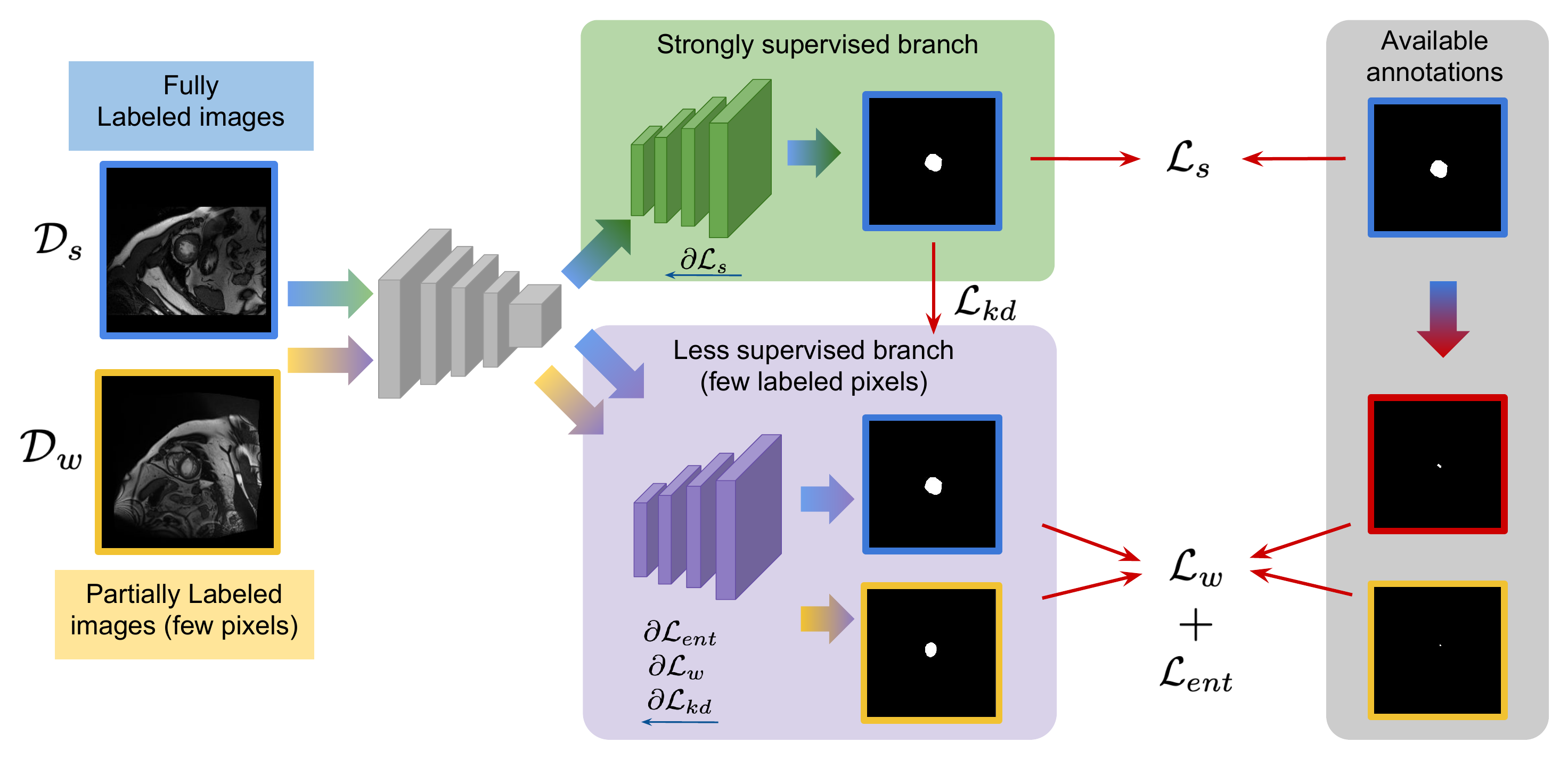}
    \caption[]{Overview of the proposed method. \extension{\textbf{Training:}} Both fully and partial labeled images are fed to the network. The top branch generates predictions for fully labeled images, whereas the bottom branch generates the outputs for partially labeled images. Furthermore, the bottom branch also generates segmentations for the fully labeled images, which are guided by the KL term between the two branches. \extension{\textbf{Inference:} Once the model is trained, we can remove the strongly supervised branch (\textit{Top}), and get the final segmentation result from the bottom stream. The gradients for each loss term are highlighted in the figure. \revB{Note that, similarly to UNet-like architectures, the proposed model has skip connections between the encoder and the decoders.}}} 
    \label{fig:WSL_strategies}
\end{figure*}

\subsection{Multi-branch architecture}

The proposed architecture is composed of multiple branches, each dedicated to a specific type of supervision (see Fig. \ref{fig:WSL_strategies}). It can be divided in two components: a shared feature extractor and independent but identical decoding networks (one per type of supervision), which  differ in the type of annotations received. \rev{It is worth mentioning that the initialisation of the weights in the decoders and the different gradients received by each branch ensure that the parameters from both decoders will not have the same values during training.} Even though the proposed multi-branch architecture has similarities with the recent work in \citep{luo2020semi}, there are significant differences, particularly in the loss functions, which leads to different optimization scenarios. 

\subsection{Supervised learning}

The top-branch is trained under the fully-supervised paradigm, where a set of training images containing pixel-level annotations for all the pixels is given, i.e., $\mathcal{D}_s$. 
The problem amounts to minimizing with respect to the network parameters a standard full-supervision loss, which typically takes the form of a cross-entropy: 
\begin{equation}
\label{eq:CE}
{\cal L}_s = - \sum_{i=1}^m \sum_{l \in \Omega_i} (\yy_i^l)^T  \log \left (\pp_i^l \right )_{\mbox{\tiny top}} 
\end{equation}
where vector $\yy_i^l = \left ( y_i^{l,1}, \dots, y_i^{l,C} \right ) \in \{0, 1\}^C$ describes the ground-truth annotation for pixel $l \in \Omega_i$. Here, notation $(\cdot)_{\mbox{\tiny top}}$ refers to the softmax outputs of the \emph{top} branch of the network. \rev{Note that all the losses are normalized by the cardinality of the training dataset, which has been omitted to simplify the notation.}

\subsection{Not so-supervised branch}

We consider the scenario where only the labels for a handful of pixels are known, i.e., scribbles or points. Particularly, we use the dataset $\mathcal{D}_w$ whose pixel-level labels are partially provided. 
Furthermore, for each image on the labeled training set, $\mathcal{D}_s$, we generate partially supervised labels (more details in the experiments' section), which are added to augment the dataset $\mathcal{D}_w$. Then, for the partially-labeled set of pixels, denoted as $\Omega_i^{\mbox{\tiny partial}}$ for each image $i \in \{1, \dots n\}$, we can resort to the following partial-supervision loss, which takes the form of a cross-entropy on the fraction of labeled pixels:
\begin{equation}
\label{eq:CE-bottom}
{\cal L}_w = - \sum_{i=m+1}^n \sum_{l \in \Omega_i^{\mbox{\tiny partial}}} (\yy_i^l)^T  \log \left (\pp_i^l \right )_{\mbox{\tiny bottom}} 
\end{equation}
where notation $(\cdot)_{\mbox{\tiny bottom}}$ refers to the softmax outputs of the \emph{bottom} branch of the network.  

\subsection{Distilling strong knowledge}

In addition to the specific supervision available at each branch, we transfer the knowledge from the teacher (top branch) to the student (bottom branch). This is done by forcing the softmax distributions from the bottom branch to mimic the probability predictions generated by the top branch for the fully labeled images in $\mathcal{D}_s$.
This knowledge-distillation regularizer takes the form of a Kullback-Leibler divergence ($\mathcal{D}_{{\mr{KL}}}$) between both distributions:
\begin{equation}
\mathcal{L}_{kd} = \sum_{i=1}^m \sum_{l \in \Omega_i}  \mathcal{D}_{{\mr{KL}}}\left ( \left (\pp_i^l \right )_{\mbox{\tiny top}} \| \left (\pp_i^l \right )_{\mbox{\tiny bottom}} \right )
\label{eq:kl}
\end{equation}
where $\mathcal{D}_{{\mr{KL}}}(\pp\|\qq)= \pp^T \log \frac{\pp}{\qq}$, with $T$ denoting the transpose operator.

\subsection{Shannon-Entropy minimization}  

Finally, we encourage high confidence in the student softmax predictions for the partially labeled images by minimizing the Shannon entropy of the predictions on the bottom branch:
\begin{equation}
\label{eq:ent-target}
\mathcal{L}_{ent} = \sum_{i=m+1}^n \sum_{l \in \Omega_i}  \mathcal{H} \left ( \pp_i^l \right )
\end{equation}
where $\mathcal{H}\left(\pp \right) = - \pp^T \log \pp$ is the Shannon entropy of distribution $\pp$.

\begin{figure}[h!]

\begin{subfigure}{.5\textwidth}
 \centering
  \includegraphics[width=1\linewidth]{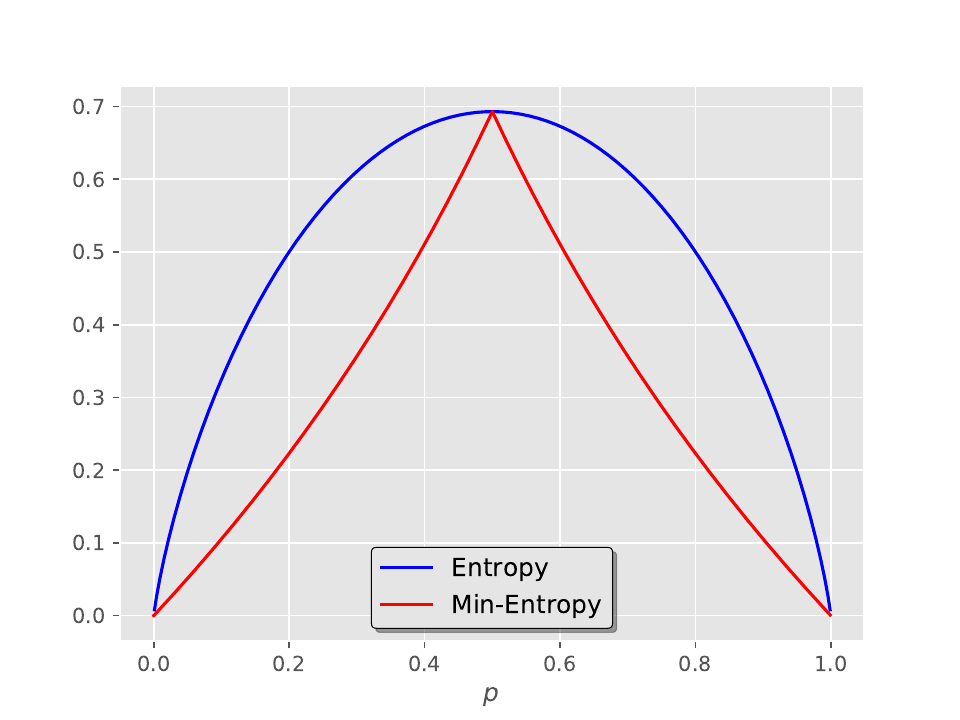}
 \caption{Shannon entropy (blue) and min-entropy (red) for a two-class distribution $(p, 1-p)$, with $p \in [0, 1]$.}
  \label{fig:entropy}
\end{subfigure}
\begin{subfigure}{.5\textwidth}
 \centering
  \includegraphics[width=1\linewidth]{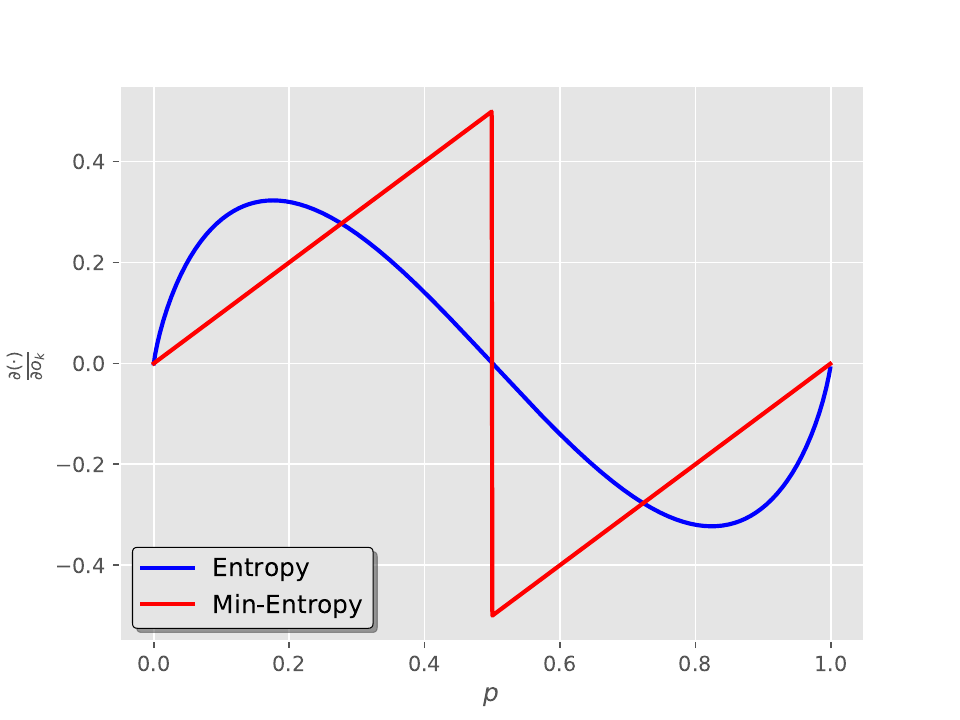}
 \caption{\rev{Derivatives of both the entropy and Min-Entropy with respect to the input logit of class $k$.}}
  \label{fig:entropy-der}
\end{subfigure}
\end{figure}

Entropy minimization is widely used in semi-supervised learning (SSL) and transductive classification \citep{grandvalet2005semi,berthelot2019mixmatch,dhillon2019baseline,boudiaf2020information} 
to encourage confident predictions at unlabeled data points. Fig. \ref{fig:entropy} plots the entropy in the case of a two-class distribution $(p, 1-p)$, showing how the minimum is reached at the vertices of the simplex, i.e., when $p=0$ or $p=1$. However, surprisingly, in segmentation, entropy is not commonly used, except a few recent works in the different contexts of SSL and domain adaptation \citep{peng2020mutual,bateson2020source,vu2019advent}. As we will see in our experiments, we found that the synergy between the entropy term for confident students, ${\cal L}_{ent}$, and the student-teacher knowledge transfer term, $\mathcal{L}_{kd}$, yields substantial increases in performances. Furthermore, in the following, we discuss an interesting link between {\em pseudo-mask generation}, common in the segmentation literature, and entropy minimization, showing that the former could be viewed as a proxy for minimizing the latter. We further provide insights as to why entropy minimization should be preferred for leveraging information from the set of unlabeled pixels. 

\subsection{Link between entropy and pseudo-mask supervision}  
\label{ssec:link}
In the weakly- and semi-supervised segmentation literature, a very dominant technique to leverage information from unlabeled pixels is to generate pseudo-masks and use these as supervision in a cross-entropy training, in an alternating way \citep{lin2016scribblesup,khoreva2017simple,papandreou2015weakly}. This self-supervision principle is also well known in classification \citep{lee2013pseudo}. Given pixel-wise predictions $\pp_i^l = (p_i^{l,1}, \dots, p_i^{l,C})$, pseudo-masks $q_i^{l,k}$ are generated as follows: $q_i^{l,k} = 1$ if $p_i^{l,k} = \max_{c} p_i^{l,c}$ and $0$ otherwise. By plugging these pseudo-labels in a cross-entropy loss, it is easy to see that this corresponds to minimizing the {\em min-entropy}, ${\mathcal H}_{\mbox{\tiny{min}}}(\pp_i^l) = - \log(\max_{c} p_i^{l,c})$, which is a lower bound on the Shannon entropy; see the red curve in Figure \ref{fig:entropy}. Fig. \ref{fig:entropy} \rev{and \ref{fig:entropy-der}} provide a good insight as to why entropy should be preferred over min-entropy (pseudo-masks) as a training loss for unlabeled data points, and our experiments confirm this. With entropy, the gradients of low-confidence predictions at the middle of the simplex are small and, therefore, dominated by the other terms at the beginning of training. However, with min-entropy, the inaccuracies resulting from uncertain predictions are reinforced (pushed towards the simplex vertices), yielding early/unrecoverable errors in the predictions, which might mislead training. This is a well-known limitation of self-supervision in the SSL literature \citep{chapelle2009semi}.

\subsection{Joint objective}

Our final loss function takes the following form:

\begin{equation}
    \mathcal{L}_{t} = \mathcal{L}_{s} + \lambda_{w} \mathcal{L}_{w} + \lambda_{kd} \mathcal{L}_{kd} + \lambda_{ent} \mathcal{L}_{ent}
    \label{eq:all_loss}
\end{equation}
where $\lambda_{w}$, $\lambda_{kd}$ and $\lambda_{ent}$ balance the importance of each term.

\section{Experimental setting}

\mypar{Benchmark dataset}
\extension{To evaluate the proposed model, we employ two public segmentation benchmarks. First, we focus} on the task of left ventricular (LV) endocardium segmentation on cine MRI images. Particularly, we used the training set from the publicly available data of the 2017 ACDC Challenge \citep{bernard2018deep}, which consists of 100 cine magnetic resonance (MR) exams covering several well defined pathologies\extension{: dilated cardiomyopathy, hypertrophic cardiomyopathy, myocardial infarction with altered left ventricular ejection fraction and abnormal right ventricle. Following prior literature, e.g., \citep{peng2020deep,Wang2020selfpaced}, slices contained in each 3D-MRI scan were considered as 2D images, whose spatial resolution was 256$\times$256.}
We split this dataset into 80 exams for training, 5 for validation and the remaining 15 for testing. \extension{Then, to demonstrate the broad applicability of our method, we use images from the Left Atrium (LA) segmentation challenge\footnote{http://atriaseg2018.cardiacatlas.org/}, which has been widely used in the context of semi-supervised segmentation. It includes 100 gadolinium-enhanced magnetic resonance imaging (GE-MRI) scans, with aligned LA segmentation masks. These 3D GE-MRI scans have isotropic resolution of $0.625\times0.625\times0.625 \text{mm}^{3}$.
Following the setting of \citep{yu2019uncertainty}, all the scans are cropped centering at the heart region for better comparison of the segmentation performance and normalized as zero mean and unit variance. Contrary to the ACDC dataset, the inputs of the network in this setting was 3D volumetric data. In our experiments, we randomly divided them into 80 for training, 5 for validation and the remaining 15 for testing.}

\begin{table*}[t!]
\scriptsize
\centering
\caption{Results on ACDC (Left-ventricle) on the testing set for the \textit{top} and \textit{bottom} branches (when applicable). Results are averaged over three runs. \rev{Best results for non-fully supervised methods are highlighted in bold.}}

\begin{tabular}{cclcc|cc|cc|cc}

\toprule
\multicolumn{2}{l}{}               &           & &    &  \multicolumn{2}{c}{\textit{Top}} & \multicolumn{2}{c}{\textit{Bottom}} & \multicolumn{2}{c}{\textit{\extension{Ensemble}}}\\
\midrule
\multicolumn{2}{c}{Setting}   &     \multicolumn{1}{c}{Model}    & \multicolumn{1}{c}{FS}    & \multicolumn{1}{c}{PS}    & DSC       & HD-95       & DSC     & HD-95  & DSC     & HD-95  \\
\midrule
\multirow{5}{*}{\textit{Set-3}}  & \multirow{4}{*}{Single Branch}&Lower bound  & $\checkmark$ & -- & 54.66 & 80.05  &  --  &   -- &  --  &   -- \\
 & &\textit{Single}  & $\checkmark$& $\checkmark$&  57.42 & 78.80  & -- &  -- &  --  &   -- \\
  & &\rev{\textit{Single + Ent}} & $\checkmark$& $\checkmark$&  \rev{43.01} & \rev{83.98}  & -- &  -- &  --  &   -- \\
& &\rev{\textit{Upper Bound (Set-3)}} & $\checkmark$& --&  \rev{87.17} & \rev{5.34}  & -- &  -- &  --  &   -- \\
 \cmidrule{2-11}
  & \multirow{3}{*}{Dual Branch}&\textit{Decoupled} \citep{luo2020semi}     & $\checkmark$ & $\checkmark$ &  56.61 & 74.95   &  5.01 & 120.06 &  --  &   -- \\
  & &\textit{Ours (KL)}  & $\checkmark$& $\checkmark$ & \extension{68.25} & \extension{63.15}  &   \extension{71.49} &  \extension{63.51}  &  --  &   --   \\
  
  & &\textit{Ours (KL+Ent)}   & $\checkmark$& $\checkmark$ &  \extension{78.38}  & \extension{46.73}  & \bf \extension{86.94}  & \bf \extension{8.84} &  \extension{86.35}  &  \extension{11.97}  \\
\midrule
\multirow{5}{*}{\textit{Set-5}}  & \multirow{4}{*}{Single Branch}&Lower bound  & $\checkmark$ & -- & 69.71&  51.75   &  --  &   --  &  --  &   --\\
 & &\textit{Single}  & $\checkmark$& $\checkmark$   & 70.73  & 51.34 & -- &  --  &  --  &   --\\
 & &\rev{\textit{Single + Ent}} & $\checkmark$& $\checkmark$&  \rev{74.92} & \rev{55.24}  & -- &  -- &  --  &   -- \\
 & &\rev{\textit{Upper Bound (Set-5)}} & $\checkmark$& --&  \rev{87.69} & \rev{4.93}  & -- &  -- &  --  &   -- \\
\cmidrule{2-11}
  & \multirow{3}{*}{Dual Branch}&\textit{Decoupled} \citep{luo2020semi}     & $\checkmark$ & $\checkmark$ & 70.96   & 54.42   &  4.29 &  127.68 &  --  &   --\\
   & &\textit{Ours (KL)}  & $\checkmark$& $\checkmark$ &  \extension{80.64} &  \extension{23.25}  &  \extension{79.06} & \extension{34.83}  &  --  &   -- \\
  
  & &\textit{Ours (KL+Ent)} & $\checkmark$& $\checkmark$  &\extension{85.57} & \extension{20.68}  &  \bf \extension{88.77} & \bf \extension{5.40} &\extension{88.54} & \extension{5.49}   \\

\midrule
\multirow{5}{*}{\textit{Set-10}} & \multirow{4}{*}{Single Branch}&Lower bound  & $\checkmark$ & -- & 78.28&  44.16    &   --  &   -- &  --  &   --\\ 
& &\textit{Single}  & $\checkmark$& $\checkmark$& 78.17  & 42.99   &  -- & --  &  --  &   --\\
 & &\rev{\textit{Single + Ent}} & $\checkmark$& $\checkmark$&  \rev{80.63} & \rev{37.75}  & -- &  -- &  --  &   -- \\
 & &\rev{\textit{Upper Bound (Set-10)}} & $\checkmark$& --&  \rev{91.18} & \rev{3.71}  & -- &  -- &  --  &   -- \\
 \cmidrule{2-11}
 & \multirow{3}{*}{Dual Branch}&\textit{Decoupled} \citep{luo2020semi}     & $\checkmark$ & $\checkmark$ &  77.53 & 32.23   &  4.58 &  125.36 &  --  &   --\\
  & &\textit{Ours (KL)}  & $\checkmark$& $\checkmark$ &\extension{88.29} &  \extension{12.47} &    \extension{88.68}  &  \extension{11.93}  &  --  &   -- \\
  & &\textit{Ours (KL+Ent)}  & $\checkmark$& $\checkmark$ & \extension{86.53} &  \extension{5.64} &   \bf \extension{90.92}  & \bf \extension{1.39} &  \extension{90.75}  &   \extension{1.55} \\
  \midrule
 All images  & Single Branch &Upper bound  & $\checkmark$ & -- & 93.31  &  3.46  &  -- & --  \\
  \bottomrule
  \multicolumn{8}{l}{\scriptsize{FS and PS indicate  \rev{fully or partially} supervised images.}}\\
\end{tabular}

\label{table:acdc-main}
\end{table*}

\mypar{Generating partially labeled images} The training exams are divided into a small set of fully labeled images, $\mathcal{D}_s$, and a larger set of images with reduced supervision, $\mathcal{D}_w$, where only a handful of pixels are labeled. Concretely, we employ the same partial labels as in \citep{kervadec2019curriculum,kervadec2019constrained}, \rev{which are obtained by eroding iteratively the full pixel-wise masks with a kernel of size 10 $\times$ 10, until the smallest possible contour is obtained.} To evaluate how increasing the amount of both fully and partially labeled affects the performance, we evaluated the proposed models in three settings, referred to as \textit{Set-3}, \textit{Set-5}, and \textit{Set-10}. In these settings, the number of fully labeled images is 3, 5 and 10, respectively, while the number of images with partial labels is $\times$5 times the number of labeled images.

\mypar{Evaluation metrics} For evaluation purposes we employ two well-known metrics in medical image segmentation: the Dice similarity score (DSC) and the modified Hausdorff-Distance (MHD). Particularly, the MHD represents the 95\textit{th} percentile of the symmetric HD between the binary objects in two images.

\mypar{Baseline methods}To demonstrate the efficiency of the proposed model, we compared it to several baselines. First, we include full-supervised baselines that will act as lower and upper bounds. The lower bound employs only a small set of fully labeled images (either 3, 5 or 10, depending on the setting), whereas the upper bound considers all the available training images. \rev{The fully labeled images used to train the lower bound baselines are exactly the same images employed in the other models.} Then, we consider a single-branch network, referred to as \textit{Single}, which receives both fully and partial labeled images without making distinction between them. \rev{In order to have a fair comparison with the proposed method, we also include a version of the \textit{Single} model where the entropy of the predictions is minimized during training, i.e., \textit{Single+Ent}.} To assess the impact of decoupling the branches without further supervision, similar to \citep{luo2020semi}, we modify the baseline network by integrating two independent decoders, while the encoder remains the same. This model, which we refer to as \textit{Decoupled}, is governed by different types of supervision at each branch \rev{and is equivalent to our model without the proposed KL and entropy minimization terms}. Then, our first model, which we refer to as \textit{KL}, integrates the KL divergence term presented in Eq. (\ref{eq:kl}), whereas \textit{KL+Ent} corresponds to the whole proposed model, which couples the two important terms in Eq. (\ref{eq:kl}) and Eq. (\ref{eq:ent-target}) in the formulation. \rev{Last, to evaluate the benefits of the proposed method compared to a equivalent fully supervised model when the same amount of images are available, we include the results when the Single model is trained with all the images on each setting, and their corresponding pixel-level mask. We refer to this model as to the \textit{Upper Bound (Set-N)}, where $N$ indicates the setting. }

\mypar{Implementation details} \extension{On the LV segmentation task}, we employed UNet \citep{ronneberger2015u} as backbone architecture for the single branch models, \extension{whereas VNet \citep{Milletari16Vnet} was utilized in the LA segmentation task. The reason behind these choices is that segmentation was performed in a 2D manner in ACDC, whereas we employ volumetric inputs for the LA dataset, following the literature.} Regarding the dual-branch architectures, we modified the decoding path of the standard UNet \extension{and VNet} to accommodate two separate branches. \rev{Note that the encoders remain the same for both single and dual-branch architectures.} All the networks \extension{on the LV segmenation task} are trained during 500 epochs by using Adam optimizer, with a batch size equal to \revB{24 (i.e., 8 labeled and 16 partially labeled images),} and the learning rate was initialized to $1 \times 10^{-4}$. We empirically set the values of $\lambda_{w}$, $\lambda_{kd}$ and $\lambda_{ent}$ to \extension{0.001}, 50 and 1, respectively. \extension{For the LA segmentation task, we followed the setting in \citep{yu2019uncertainty} by training the network with SGD optimizer and batch size equal to \revB{12 (i.e., 4 labeled and 8 partially labeled images),}.} We found that our formulation provided the best results when the input distributions to the KL term in eq. (\ref{eq:kl}) were very smooth, which was achieved by applying softmax over the softmax predictions. All the hyperparameters, \rev{for all the baselines and models}, were fixed by using the independent validation set. \rev{In particular, we first found the best value for $\lambda_{w}$ (0.001) for the \textit{Single} model, which was then fixed to find the other hyperparameters of the proposed model. Then, the optimal $\lambda_{kd}$ weight of the KL term was found (Detailed results can be found in Supplemental Materials, section \ref{sec:paramsearch}). Last, with $\lambda_{w}$ and $\lambda_{kd}$ fixed, we found the best value of $\lambda_{ent}$, whose value is the same in both the proposed model and the baseline \textit{Single+Ent}.} Furthermore, we perform 3 runs for each model and report the average values. The code was implemented in PyTorch and all the experiments were performed in a server equipped with a NVIDIA Titan RTX GPU.

\subsection{Results}

\paragraph{\textbf{Main results}} Table \ref{table:acdc-main} reports the quantitative evaluation of the proposed method compared to the different baselines \extension{on the ACDC dataset}. 

\extension{The first thing we observe is that, across all the settings,} simply adding partial annotations to the training set does not considerably improve the segmentation performance, \extension{i.e., \textit{Single} model}. \rev{Furthermore, integrating the entropy minimization term in this baseline results in a performance degradation in the less-supervised setting (i.e., Set-3), whereas the performance typically increases in the other two settings. Indeed, the lower performance observed in the setting Set-3 might be due to the entropy minimization term pushing towards trivial solutions. This is particularly important on this setting: as the number of fully-labeled images is low the information derived from these images might not be enough to serve as a strong prior to avoid these trivial solutions.}

On the other hand, by integrating the guidance from the upper branch, the network is capable of leveraging additional partially-labeled images more efficiently through the bottom branch. Furthermore, if we couple the KL divergence term with an objective based on minimizing the entropy of the predictions on the partial labeled images, the segmentation performance substantially increases. Particularly, the gain obtained by the complete model is consistent across the several settings, improving the DSC by 6-12\% compared to the \textit{KL} model, and reducing the MHD by nearly 30\%. Compared to the baseline dual-branch model, i.e., \textit{Decoupled}, our approach brings improvements of 10-20\% in terms of DSC and reduces the MHD values by 30-40\%. \rev{Last, the results obtained by the proposed model in each setting are on par with their individual upperbound counterparts (i.e., those trained with the same images as our model), particularly in terms of DSC. Furthermore, as the number of labeled images increases, the gap in the HD metric is decreased between the two models, with our model outperforming the upper bound in the \textit{Set-10} setting.}

These results demonstrate the strong capabilities of the proposed model to leverage fully and partially labeled images during training. It is noteworthy to mention that findings on these results, where \textbf{the student excels the teacher}, align with recent observations in classification \citep{furlanello2018born,yim2017gift}. \extension{Note that the top branch is also improved in our formulation. This can be explained from the fact that even though the supervision the teacher (\textit{top}) receives remains unchanged, changes in the student (\textit{bottom}) also affect the encoder, which is shared among both. Thus, the integration of the proposed objective also results in an improvement on the latent representation of the model.}

\extension{The predictions from top and bottom streams can be seen as independent model outputs. In this scenario, ensemble learning has often demonstrated to be an efficient solution to boost the performance of single models \citep{dolz2020deep}. Nevertheless, as only two different predictions are available, and there exists a significant gap in performance between both, combining both outputs hampers the performance compared to the bottom branch, particularly in lower data regimes. This supports our choice of using only the weakly supervised stream at inference. }

\paragraph{\textbf{Comparison with proposals}}

As mentioned previously, a popular paradigm in weakly and semi-supervised segmentation is to resort to pseudo-masks generated by a trained model, which are used to re-train the network mimicking full supervision. 
To demonstrate that our model leverages more efficiently the available data, we train a network with \revB{an augmented dataset composed by the available labeled images and the} proposals generated \revB{on the unlabeled images} by the \textit{Lower bound} and \textit{KL} models, whose results are reported in Table \ref{table:acdc-kl}. We can observe that despite typically improving the base model, minimizing the cross-entropy over proposals does not outperform directly minimizing the entropy on the predictions of the partially labeled images. 

\begin{table}[h!]
\scriptsize

\centering
\caption{Results obtained by training on an augmented dataset composed by fully labeled images and proposals generated from the \textit{Lower bound} and \textit{KL} models \extension{on the test set}. \revB{Results on the ACDC dataset}.} 
\resizebox{0.9\columnwidth}{!}{
\begin{tabular}{l|cc|cc|cc}

\toprule

\multicolumn{1}{l}{} & \multicolumn{2}{c}{\textit{\begin{tabular}[c]{@{}c@{}}Proposals\\ (Lower bound)\end{tabular} }} & \multicolumn{2}{c}{\textit{\begin{tabular}[c]{@{}c@{}}Proposals\\ (KL)\end{tabular} }} & \multicolumn{2}{c}{\textit{\begin{tabular}[c]{@{}c@{}}Ours\\ (KL+Ent)\end{tabular} }} \\
\midrule
\multicolumn{1}{l}{Setting}&  DSC       & HD-95  & DSC       & HD-95   & DSC       & HD-95          \\
\midrule
 \textit{Set-3}    & 63.11 & 49.99 & \extension{74.27} &  \extension{45.44} & 
 \bf \extension{86.94} & \bf \extension{8.84}  \\
 \textit{Set-5}    & 73.91 & 45.54& \extension{81.35} &  \extension{20.28} & 
 \bf \extension{88.77} & \bf \extension{5.40}  \\
 \textit{Set-10}    & 81.31 & 29.95 & \extension{89.26}  & \extension{7.98}  & 
 \bf \extension{90.92} & \bf \extension{1.39}  \\
  \bottomrule
\end{tabular}
\label{table:acdc-kl}
}
\end{table}

\begin{figure}[h!]
    \centering
    \includegraphics[width=0.5\textwidth]{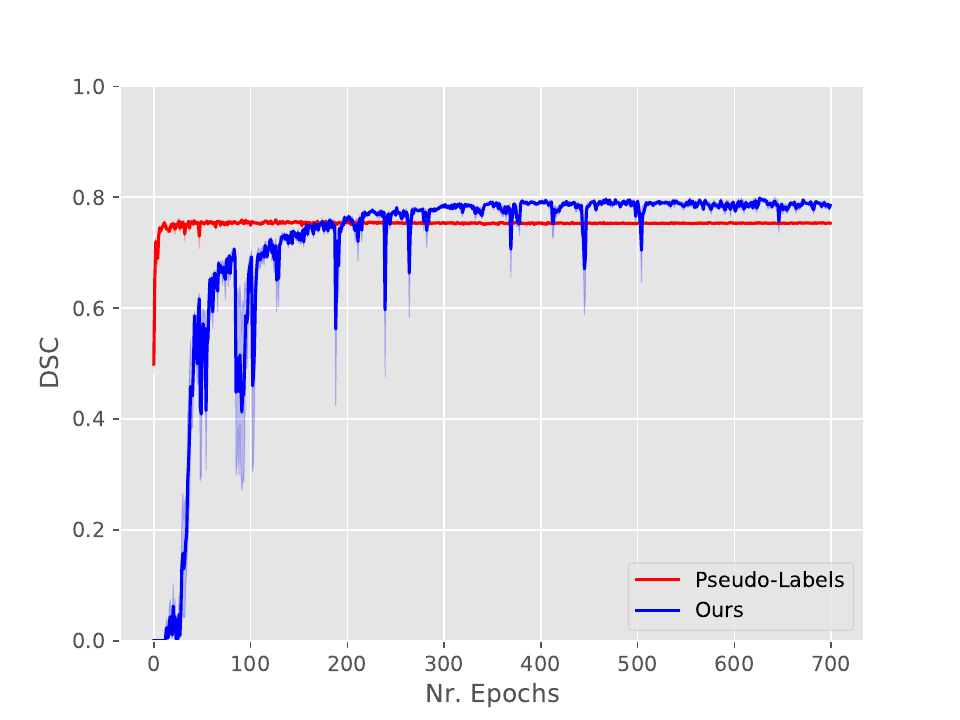}
    \caption[]{\extension{DSC evolution across the training when trained with pseudo-labels (\textit{red}) and the proposed formulation (\textit{blue}) on 3 subjects fully labeled on the ACDC validation dataset.}} 
    \label{fig:pseudo_vs_ours}
\end{figure}

\extension{In addition to the quantitative results reported below, we depict the performance evolution on the validation set, for the setting Set-3, in Fig. \ref{fig:pseudo_vs_ours}. An interesting observation is that, while training with pseudo-labels converges faster, our method needs more iterations to reach convergence which is achieved at a slower pace. Nevertheless, minimizing the entropy on the predictions results in the model outperforming the pseudo-label based approach. We argue that this behaviour can be explained from a gradient dynamics perspective. At the beginning of the training, there exist low-confidence predictions which might result in inaccurate predictions (e.g., in Fig \ref{fig:entropy} we can observe that these predictions lie within the middle of the simplex). As we showed in Section \ref{ssec:link}, employing pseudo-masks in a cross-entropy loss corresponds to minimizing the \textit{min-entropy}, which quickly pushes low-confidence predictions towards the simplex vertices at the beginning of the training. On the other hand, if we employ an entropy term, the gradients of low-confidence predictions in the same region (i.e., middle of the simplex) are small compared to the other terms at the beginning. However, as the other terms start to be satisfied, the scale of their gradients becomes comparable to the entropy gradients term, and hence this term begins its regularization role.}

\begin{figure}[h!]
    \centering
    \includegraphics[width=0.475\textwidth]{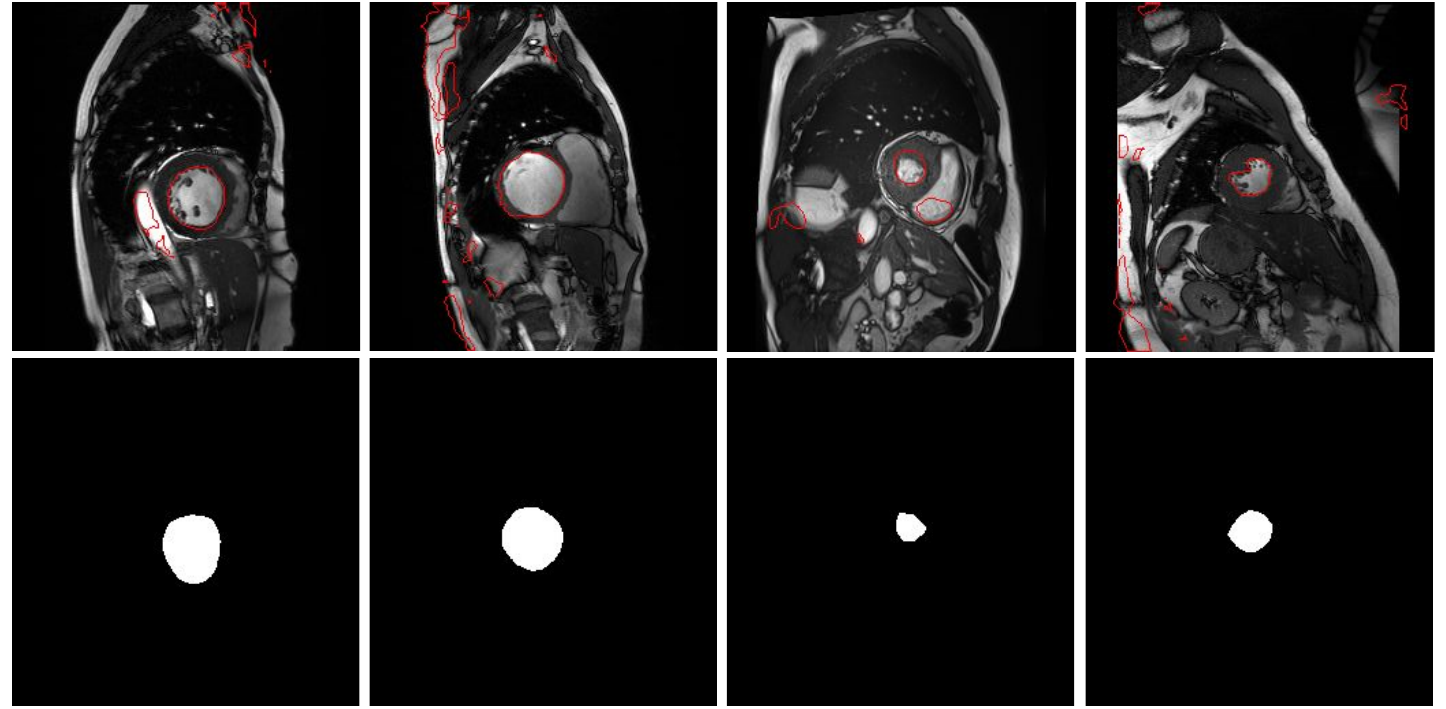}
    \caption[]{\extension{Failure cases which are employed as pseudo-labels in the \textit{proposals-based} approach \rev{(top)}, whose errors are reinforced during training, \rev{and their corresponding ground truth}. Best viewed in colours.}}
    \label{fig:failure_Proposals}
\end{figure}

\extension{Several failure cases from proposals are depicted in Fig \ref{fig:failure_Proposals}. With this strategy, these errors are propagated during training, which might explain the low performance compared to our method. In addition to lower performances, this approach requires to fully train a first model, generate pseudo-labels and then re-train a second model with the generated masks. This contrasts to our method, which is trained in an end-to-end manner.}

\paragraph{\extension{\textbf{Are errors actually propagated through the training?}}}

\begin{figure}[h!]
    \centering
    \includegraphics[width=0.475\textwidth]{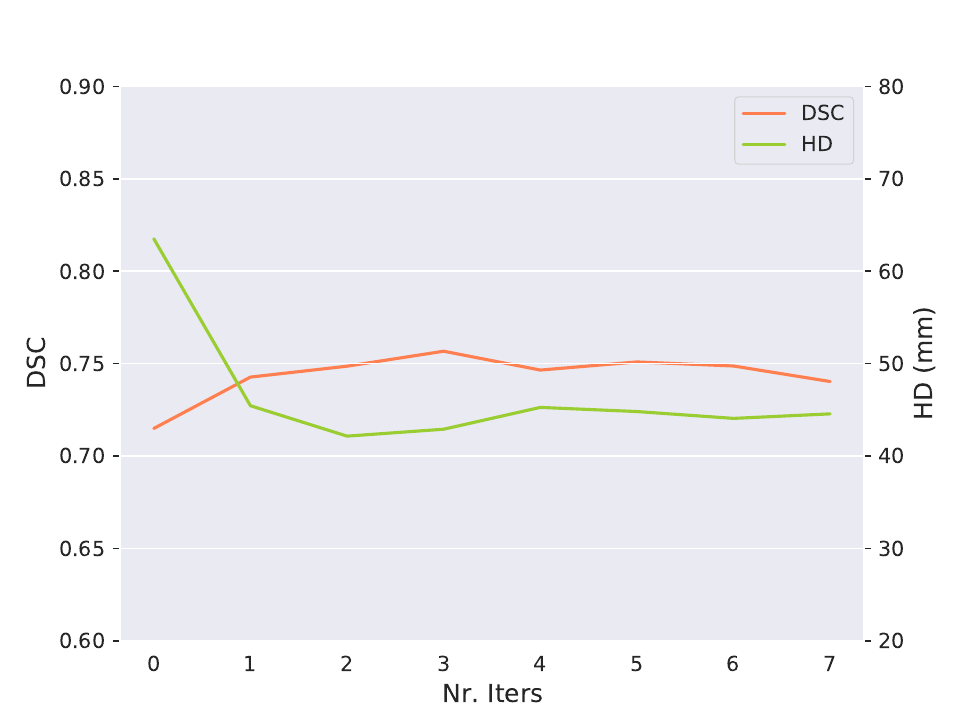}
    \caption[]{\extension{Evolution of the results from the \textit{pseudo-labels} after several iterations (\rev{on the test set of ACDC for the Setting-3}).}}
    \label{fig:pseudo-collapse}
\end{figure}

\extension{To illustrate the weaknesses of pseudo labels based approaches, particularly in reinforcing the errors, we perform the following experiment. We start with an initial model, i.e., \textit{Ours (KL)}, generates the pseudo labels to train the model at iteration $\mathcal{I}_1$, i.e., \textit{Proposals (KL)}. Similarly, once the model at iteration $\mathcal{I}_t$ is trained, it is used to generate the pseudo-labels for training the next model at iteration $\mathcal{I}_{t+1}$. Furthermore, at each iteration, the model parameters are initialized randomly (note that this is similar to the so-called self-training strategy). We see in Figure \ref{fig:pseudo-collapse} that, despite having a performance improvement in early iterations, 
this improvement quickly saturates. This degradation of results over time suggests that the model is unable to correct noisy pseudo-labels and accumulates these errors across iterations. As explained recently in \citep{huo2021atso}, this \textit{trapping} effect can be explained from an optimization perspective. If we use $\XX_i$ to denote a training image and $\YY_i$ its corresponding ground truth, we can assume that the predicted segmentation is $\hat{\YY}_i=\YY_i+\ee_i$, where $\ee_i$ denotes the prediction error. If $\XX_i$ belongs to the fully labeled dataset $\mathcal{D}_s$, $\YY_i$ is known. Thus, as the optimization objective involves minimizing $|\ee_i|$ on $\mathcal{D}_s$, it follows a zero mean distribution. In contrast, $\YY_i$ is unknown on the weakly supervised dataset $\mathcal{D}_w$, which allows $\ee_i$ to follow a distribution with non-zero mean. In this scenario, the prediction $\hat{\YY}_i$ (used later as pseudo-label) might integrate this noise, which can be propagated to the model in subsequent iterations.}

\paragraph{\extension{\textbf{Comparison with semi-supervised methods}}}

\extension{We now compare the proposed approach with a series of state-of-the-art semi-supervised segmentation approaches including: UA-MT \citep{yu2019uncertainty}, GLMI \citep{Peng2021boost} and SSCO \citep{Wang2020selfpaced}. UA-MT is an established method to benchmark semi-supervised segmentation approaches, whereas GLMI and SSCO have recently demonstrated superior performance over the existing literature.
\rev{We carefully tune the hyper-parameters and report the best performance for each method. 
For each semi-supervised method, we fixed the number of labeled and unlabeled training examples in each mini-batch to 4 as this yield consistent results.
For UA-MT, the consistency weight is set to 0.1 and increased by a Gaussian ramp-up function during the first 100 epochs. 
For GLMI \citep{Peng2021boost}, the three balancing weights for controlling the relative contributions of global mutual information (MI), local MI and consistency loss are set to 1.0, 0.1 and 10.0, respectively. To evaluate SSCO, the light-weight architecture ENet used in the original implementation \citep{Wang2020selfpaced} is replaced by UNet for a fair comparison in our experiments. The values for the hyper-parameters are the same as those reported in \citep{Wang2020selfpaced}, as it empirically worked well in our experiment. For detailed parameter search for each method, please refer to Section \ref{sec:paramsearch-semi} in the Supplemental materials.}
Table \ref{table:ssl} reports the results from this study. We can observe that, under the same conditions, our approach significantly outperforms SSL state-of-the-art methods, particularly in the scenarios where less labeled images are available. For example, when only 3 scans are fully labeled, our method significantly outperforms the recent method in \citep{Wang2020selfpaced}, with a gap of nearly 10$\%$ in terms of DSC. Even though one can argue that our method employs more supervision than these approaches, the cost of it is negligible, as extra labeled pixels could mimic the human behaviour of quickly drawing scribbles in a volumetric scan.  }

\begin{table}[h!]
\scriptsize
\centering
\caption{\extension{Comparison to semi-supervised segmentation approaches on the ACDC test dataset. Results are averaged over three runs.}}

\begin{tabular}{cl|cc}

\toprule
Setting &     \multicolumn{1}{c}{Model}    &  DSC       & HD-95       \\
\midrule
\multirow{5}{*}{\textit{Set-3}}  & Lower bound  &  54.66 & 80.05   \\
  & \extension{UA-MT \citep{yu2019uncertainty}} *  &  \extension{70.62} & \extension{39.06}  \\
 & \extension{GLMI \citep{Peng2021boost}} *  &  \extension{76.27} & \extension{11.21}  \\
 & \extension{SSCO \citep{Wang2020selfpaced}} * &  \extension{77.16} & \extension{10.10} \\
  & \textit{Ours (KL+Ent)*}   &   \bf \extension{86.94}  & \bf \extension{8.84}   \\
\midrule
\multirow{5}{*}{\textit{Set-5}}  & Lower bound  & 69.71&  51.75     \\
& \extension{UA-MT \citep{yu2019uncertainty}} *  &  \extension{74.71} & \extension{30.36}  \\
 & \extension{GLMI \citep{Peng2021boost}} *  &  \extension{80.58} & \extension{8.14}  \\
  & \extension{SSCO \citep{Wang2020selfpaced}} * &  \extension{81.17} & \extension{8.15} \\

  & \textit{Ours (KL+Ent)} &   \bf \extension{88.77} & \bf \extension{5.40}   \\
\midrule
\multirow{5}{*}{\textit{Set-10}} & Lower bound  &  78.28&  44.16   \\ 
& \extension{UA-MT} \citep{yu2019uncertainty} *  &  \extension{82.67} & \extension{28.04}  \\
 & \extension{GLMI \citep{Peng2021boost}} *  &  \extension{88.43} & \extension{8.04}  \\
 & \extension{SSCO \citep{Wang2020selfpaced}} * &  \extension{89.25} & \extension{2.58} \\
  & \textit{Ours (KL+Ent)}  &   \bf \extension{90.92}  & \bf \extension{1.39}  \\

  \bottomrule
\multicolumn{3}{l}{\scriptsize{*Ours uses partially supervised images.}}\\
\end{tabular}
\label{table:ssl}
\end{table}

\paragraph{\textbf{Ablation study on the importance of the KL term}}

The objective of this ablation study (Table \ref{tab:results-acdc-ablationK}) is to assess the effect of balancing the importance of the KL term in our formulation. Particularly, the KL term plays a crucial role in the proposed formulation, as it guides the entropy term during training to avoid degenerate solutions. We note that the value of the KL term is typically 2 orders of magnitude smaller than the entropy objective. Therefore, by setting its weight ($\lambda_{{\mr{KL}}}$) to 1, we demonstrate empirically its crucial role during training when coupled with the entropy term, as in this setting the latter strongly dominates the training. In this scenario, we observe that the model is negatively impacted, particularly when fully-labeled images are scarce, i.e., \textit{Set-3}, significantly outperforming the lower bound model. This confirms our hypothesis that minimizing the entropy alone results in degenerated solutions. Increasing the weight of the KL term typically alleviates this issue. However, if much importance is given to this objective the performance also degrades. This is likely due to the fact that the bottom branch is strongly encouraged to follow the behaviour of the top branch, and the effect of the entropy  term is diminished.

\begin{table}[t!]
\scriptsize
\centering
\caption{\extension{Impact of $\lambda_{{\mr{KL}}}$ on the proposed formulation. \revB{Results on the ACDC dataset}.
}}
\resizebox{0.9\columnwidth}{!}
{
\begin{tabular}{l|cc|cc|cc|}
\toprule
         & \multicolumn{2}{c}{\textit{Set-3}} & \multicolumn{2}{c}{\textit{Set-5}} & \multicolumn{2}{c}{\textit{Set-10}} \\
         \midrule
         & DSC        & HD-95       & DSC        & HD-95       & DSC        & HD-95        \\
\midrule
$\lambda_{K}=0.1$  & 66.36 & 51.86  & 77.26 & 31.30 & 81.95 & 30.11 \\
$\lambda_{K}=1$  & 71.31 & 39.67  & 83.88 & 21.68 & 89.73 & 7.34 \\
$\lambda_{K}=10$  & 85.87  & 12.38  & 86.52 & 7.86& 90.55 & 2.84\\
$\lambda_{K}=50$    & \bf 86.94  & \bf 8.84  & \bf 88.77 & \bf 5.40& \bf 90.92 & \bf 1.39\\
$\lambda_{K}=100$   & 83.92  & 18.17  & 87.31 & 9.34& 89.38 & 1.62\\
$\lambda_{K}=1000$   & 76.20  & 29.99  & 85.59 & 13.93& 88.90 & 4.46\\   
\bottomrule
\end{tabular}
}
\label{tab:results-acdc-ablationK}
\end{table}

\paragraph{\extension{\textbf{Sensitivity to $\lambda_{w}$}}}\extension{This ablation study quantifies the contribution of the partially labeled cross-entropy term in Eq. (\ref{eq:all_loss}) by evaluating the performance across several $\lambda_{w}$ values. In this study, the number of \rev{partially labeled} images is 5 times larger than the number of \rev{fully} labeled images. Table \ref{tab:results-acdc-ablation_lambdaW} reports these results, from which we can see that a value of $\lambda_{w}=0.001$ consistently brings the best performance across all the settings. Differences are larger in the Set-3 case, which corresponds to the lowest amount of extra information, in terms of partial labels. We believe that, as the amount of additional supervision increases, the performance is less sensitive to the weight of this term.}

\begin{table}[h!]
\scriptsize
\centering
\caption{\extension{Impact of $\lambda_{w}$ on the proposed formulation. \revB{Results on the ACDC dataset}.}}
\resizebox{0.9\columnwidth}{!}
{
\begin{tabular}{l|cc|cc|cc|}
\toprule
         & \multicolumn{2}{c}{\textit{Set-3}} & \multicolumn{2}{c}{\textit{Set-5}} & \multicolumn{2}{c}{\textit{Set-10}} \\
         \midrule
         & DSC        & HD-95       & DSC        & HD-95       & DSC        & HD-95        \\
\midrule
\extension{$\lambda_{w}=1$}  & \extension{80.55} & \extension{20.35}  & \extension{84.09}  & \extension{18.78}  & \extension{90.26}  & \extension{6.99}  \\
\extension{$\lambda_{w}=0.1$}  & \extension{84.09}  &  \extension{11.46} &  \extension{87.09} & \extension{18.03} & \extension{90.17} & \extension{4.84}\\
\extension{$\lambda_{w}=0.01$}  & \extension{85.40}  & \extension{12.05} & \extension{87.39} & \extension{9.87} & \extension{90.14} & \extension{2.70} \\
\extension{$\lambda_{w}=0.001$}    & \extension{\bf 86.94}  & \extension{\bf 8.84} & \extension{\bf 88.77}  & \extension{\bf 5.40} & \extension{\bf 90.92} &  \extension{\bf 1.39}     \\
\extension{$\lambda_{w}=0.0001$}   & \extension{81.24}  &  \extension{19.77} & \extension{83.68} & \extension{15.88} & \extension{87.64} & \extension{4.39} \\    
\bottomrule
\end{tabular}
}
\label{tab:results-acdc-ablation_lambdaW}
\end{table}

\extension{\mypar{On the divergence terms.}} \extension{In addition to the widely well-known KL-divergence, we study a series of additional divergences for the constraining term in Eq. (\ref{eq:kl}). In particular, we first resort to the Bhattacharyya distance \citep{bhattacharyya1946some}. For two discrete distributions $\pp=(p_k)^K_{k=1}$ and $\qq=(q_k)^K_{k=1}$ this term takes the following form:
\begin{equation}
\mathcal{D}_{BC}(\pp,\qq)= - \log \sum_{k=1}^K \left ( p_k q_k \right )^{\frac{1}{2}} 
\label{eq:bc}
\end{equation}
Furthermore, we also investigate the Tsallis's formulation of $\alpha$-divergence \citep{cichocki2010families,tsallis1988possible}, which generalizes the KL:
\begin{align}
\mathcal{D}_{\alpha}(\pp\|\qq) &= -\sum_{k=1}^K p_k \log_{\alpha} \left ( \frac{q_k}{p_k} \right)\nonumber\\
&= \frac{1}{1-\alpha}  \left ( 1 - \sum_{k=1}^K p_k^{\alpha} q_k ^{1-\alpha}\right ) 
\label{eq:alphadiv}
\end{align}

Table \ref{table:table_diver} reports the results generated by the different divergence functionals, evaluated on the \textit{Set-3} setting. We see that there are two cases, i.e., Bhattacharyya distance and $\alpha$-divergence with $\alpha\!=\!2.0$, that outperforms the model integrating the KL-divergence. This suggests that our model can be further improved by replacing the KL term by alternative divergence functions as consistency losses.}

\begin{table}[h!]
\scriptsize
\centering
\caption{\extension{Comparison of several divergence functions for the third term (i.e., $\mathcal{L}_{kd}$ term in eq. \ref{eq:all_loss}), on the \textit{Set-3} setting. \revB{Results on the ACDC dataset}.}} 

\begin{tabular}{cl|cc}

\toprule
Setting &     \multicolumn{1}{c}{Model}    &  DSC       & HD-95       \\
\midrule
\multirow{5}{*}{\textit{Set-3}}  & Kullback-Leibler  &  \extension{86.94} & \extension{8.84}   \\
 & Bhattacharyya  &  \extension{86.98}  & \bf \extension{5.26}  \\

  & $\alpha-$Divergence ($\alpha=2.0$)  & \extension{\bf 88.04}  & \extension{6.43}    \\
  & $\alpha-$Divergence ($\alpha=3.0$)  & \extension{86.89}  &  \extension{10.24}  \\
  & $\alpha-$Divergence ($\alpha=5.0$)  & \extension{85.57}  &     \extension{11.77}\\
  \bottomrule
\end{tabular}
\label{table:table_diver}
\end{table}

\begin{figure}[h!]
    \centering
    \includegraphics[width=0.5\textwidth]{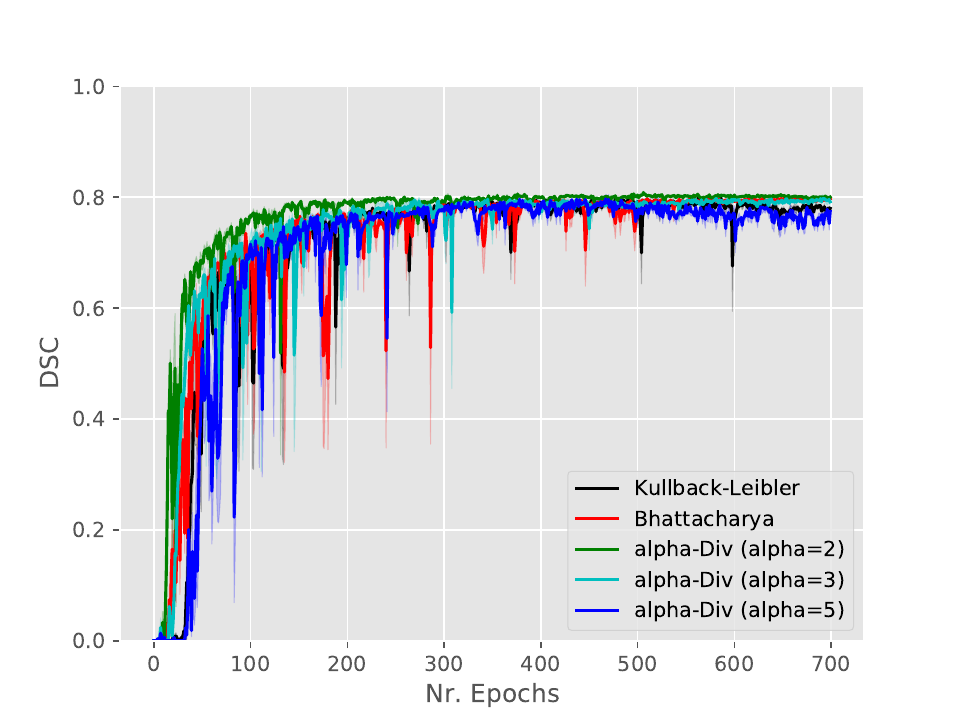}
    \caption[]{\extension{Comparison of different divergences in terms of DSC in the validation set (on the \textit{Set-3} setting).}}
    \label{fig:divergs}
\end{figure}

\extension{Figure \ref{fig:divergs} depicts the evolution, in terms of DSC, on the validation set for the different divergences. Despite showing a similar performance, employing the $\alpha$-divergence with $\alpha\!=\!2$ (green line) stands out from the others, with a faster convergence at the beginning of the training.}  

\begin{figure}[h!]
\centering
 \includegraphics[width=0.9\linewidth]{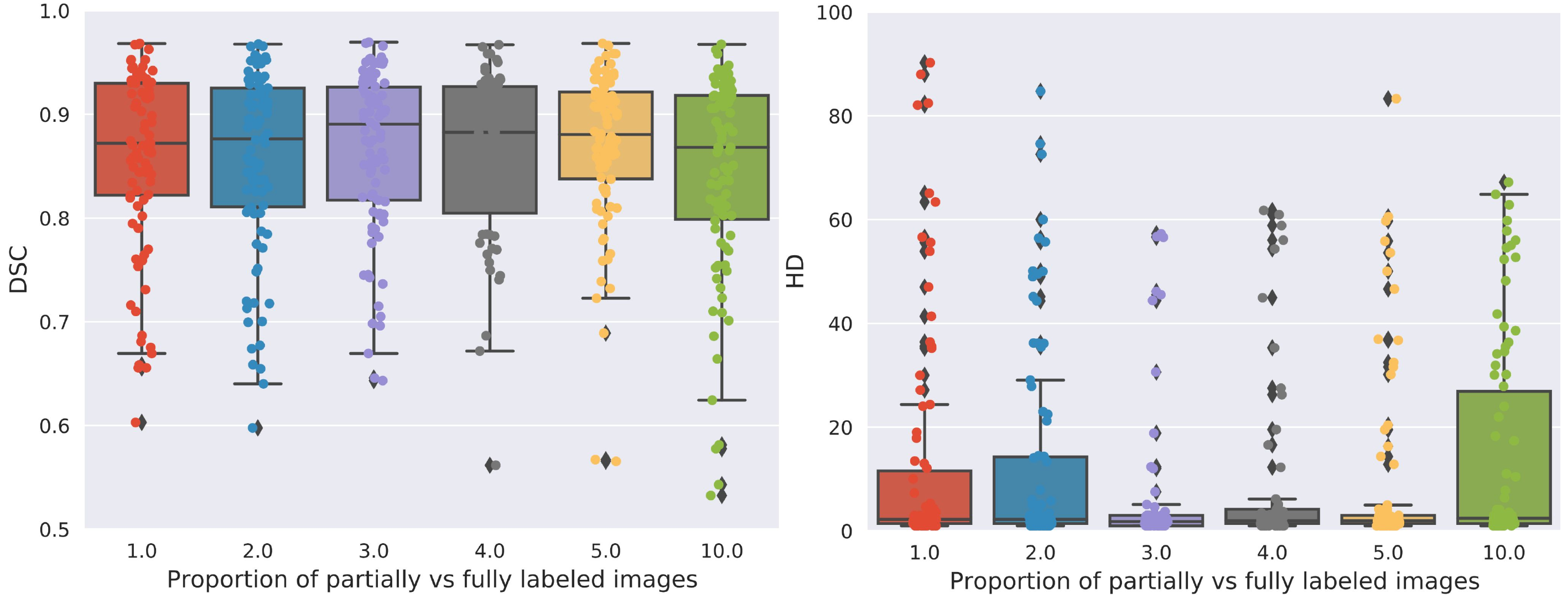}
 \caption{\extension{Ablation experiments on the effect of increasing the number of partially labeled images for a fix set of labeled images (on the \textit{Set-3} setting for ACDC). The value in the x axis represents the amount of partially labeled images with respect the labeled images, e.g., 2.0 indicates that there are 2 times more partially labeled than fully labeled images.}}
\label{fig:unlab}
\end{figure}

\begin{figure*}[t!]
    \centering
    \includegraphics[width=0.9\textwidth]{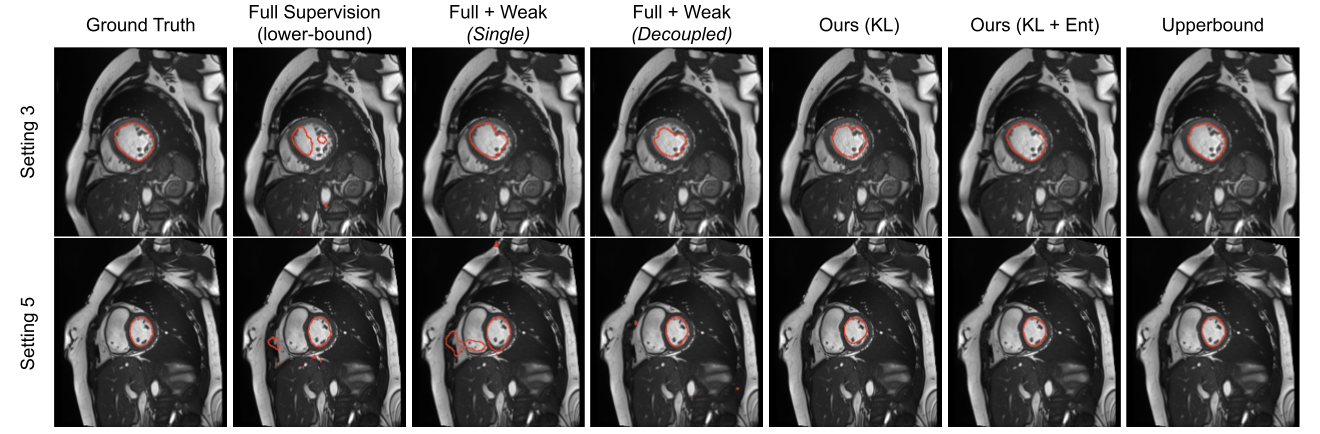}
    \caption[]{Qualitative results for the analyzed models under two different settings.}
    \label{fig:qualitative-main}
\end{figure*}

\begin{figure}[h!]
    \centering
    \includegraphics[width=0.5\textwidth]{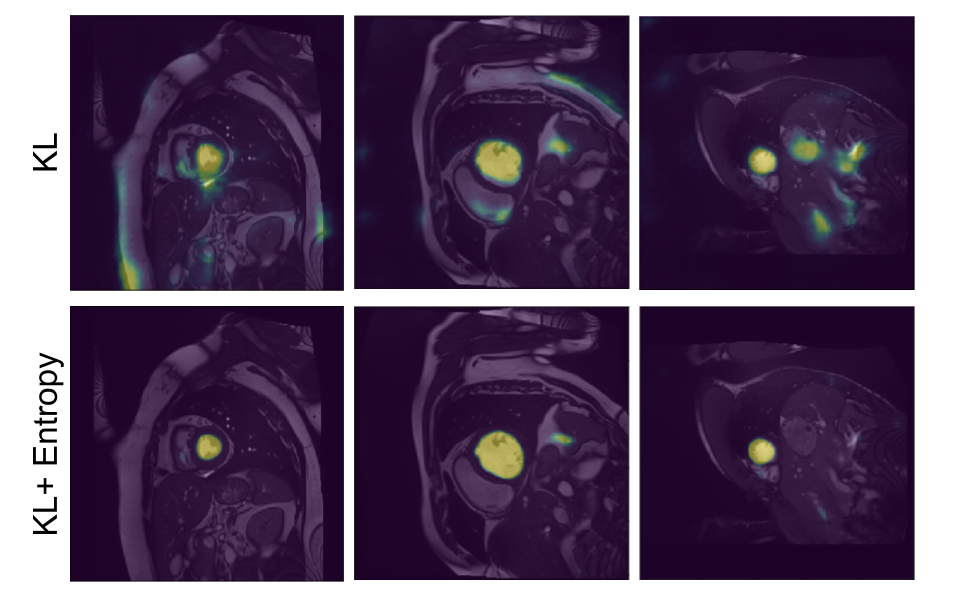}
    \caption{Probability maps obtained by the proposed $KL$ and $KL+Ent$ models.}
    \label{fig:KL-vs-Entropy}
\end{figure}

\begin{figure*}[htb]
    \centering
    \includegraphics[width=1\textwidth]{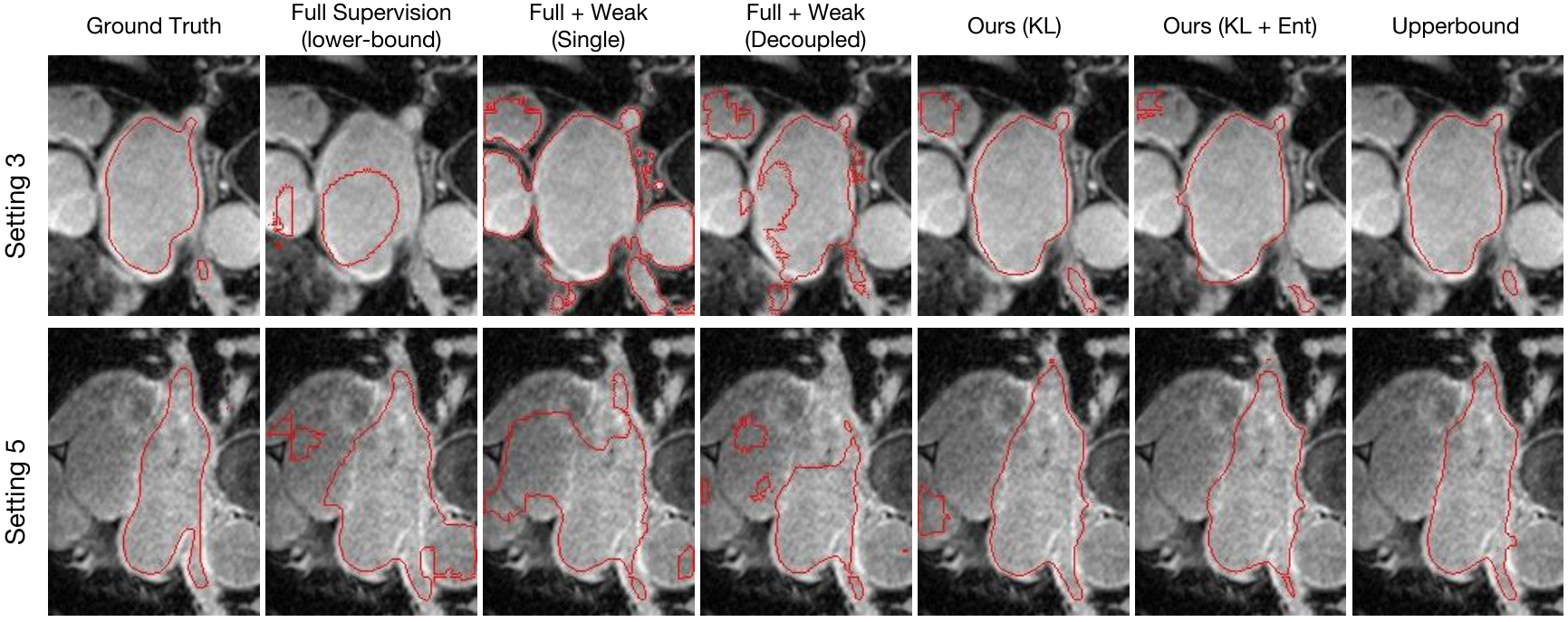}
    \caption[]{\extension{Qualitative results on LA segmentation for the analyzed models under two different settings.}}
    \label{fig:qualitative-la}
\end{figure*}

\paragraph{\extension{\textbf{Impact on the number of \rev{partially labeled} images}}}
\extension{We now evaluate the impact of training the proposed model with a diverse amount of \rev{partially labeled} images. These results, which are depicted in Fig \ref{fig:unlab}, show that by having a ratio between labeled and \rev{partially labeled} data ranging from 3 to 5 typically brings the best performance, both in terms of DSC and HD distance. It is important to note that these results represent the lowest supervised scenario, where only 3 fully labeled images are available. Nevertheless, as seen in previous ablation, as the number of fully labeled cases becomes larger, the impact of several elements is reduced.  }

\paragraph{Qualitative results} In addition to the numerical results presented before, we also depict qualitative results in Fig. \ref{fig:qualitative-main} and Fig. \ref{fig:KL-vs-Entropy}. Particularly, Fig. \ref{fig:qualitative-main} depicts the segmentation results for the models evaluated in Table \ref{table:acdc-main}. We see that results obtained by models with a single network typically under-segment the object of interest (\textit{first row}) or generate many false positives (\textit{second row}). Decoupling the decoding branches might reduce the false positive rate, however, it also tends to under-segment the target. Finally, we observe that both of our formulations achieve qualitatively better segmentation results, with the \textit{KL+Ent} model yielding segmentations similar to those generated by the upper bound model. Furthermore, in Fig. \ref{fig:KL-vs-Entropy}, we illustrate additional qualitative results of our models. We observe that without the entropy term our model produces less confident predictions, which results in more noisy segmentations.

\begin{table*}[h!]
\scriptsize
\centering
\caption{\extension{Results on the test set of LA segmentation for the \textit{top} and \textit{bottom} branches (when applicable). Results are averaged over three runs. Best results highlighted in bold and second best results are underlined.}}
\resizebox{1.0\textwidth}{!}{
\begin{tabular}{cclcc|cc|cc|cc}

\toprule
\multicolumn{2}{l}{}               &           & &    &  \multicolumn{2}{c}{\textit{Top}} & \multicolumn{2}{c}{\textit{Bottom}} & \multicolumn{2}{c}{\textit{Ensemble}} \\
\midrule
\multicolumn{2}{c}{Setting} &     \multicolumn{1}{c}{Model}    & \multicolumn{1}{c}{FS}    & \multicolumn{1}{c}{PS}    & DSC       & HD-95       & DSC     & HD-95 & DSC     & HD-95 \\
\midrule
\multirow{5}{*}{\textit{Set-3}} & \multirow{4}{*}{Single Branch} & Lower bound  & $\checkmark$ & -- & \extension{37.76} & \extension{42.30}  &  --  &   --  \\
& & \textit{Single}  & $\checkmark$& $\checkmark$&  \extension{38.11}  &  \extension{34.28} &  --  &   -- &  --  &   --   \\
& &\rev{\textit{Single + Ent}} & $\checkmark$& $\checkmark$&  \rev{41.82} & \rev{36.54}  & -- &  -- &  --  &   -- \\
& &\rev{\textit{Upper Bound (Set-3)}} & $\checkmark$& --&  \rev{85.91} & \rev{15.10}  & -- &  -- &  --  &   -- \\
\cmidrule{2-11}
 & \multirow{3}{*}{Dual Branch} & \textit{Decoupled} \citep{luo2020semi}     & $\checkmark$ & $\checkmark$ & \extension{57.85}  & \extension{42.04}   & \extension{18.11}  & \extension{61.53} &  --  &   --  \\
  & & \textit{Ours (KL)}  & $\checkmark$& $\checkmark$ & \extension{61.41} & \extension{33.63}  &   \extension{66.75} &  \extension{33.71} &  --  &   -- \\
  & & \textit{Ours (KL+Ent)}   & $\checkmark$& $\checkmark$ &  \extension{\underline{69.94}}  & \extension{\underline{33.05}}  & \extension{\textbf{72.09}}  & \extension{\textbf{32.33}} &  \extension{71.29}  &   \extension{32.16} \\
\midrule
\multirow{5}{*}{\textit{Set-5}} & \multirow{4}{*}{Single Branch} & Lower bound  & $\checkmark$ & -- & \extension{64.86} & \extension{35.97}   &  --  &   --  \\
 & & \textit{Single}  & $\checkmark$& $\checkmark$   &  \extension{72.06}  & \extension{28.97}  & -- &  -- &  --  &   --  \\
 & &\rev{\textit{Single + Ent}} & $\checkmark$& $\checkmark$&  \rev{71.79} & \rev{28.62}  & -- &  -- &  --  &   -- \\
 & &\rev{\textit{Upper Bound (Set-5)}} & $\checkmark$& --&  \rev{86.21} & \rev{14.01}  & -- &  -- &  --  &   -- \\
 \cmidrule{2-11}
 & \multirow{3}{*}{Dual Branch} & \textit{Decoupled} \citep{luo2020semi}     & $\checkmark$ & $\checkmark$ &  \extension{75.33}  & \extension{29.06} &  \extension{17.58}  & \extension{61.18} &  --  &   -- \\
  & & \textit{Ours (KL)}  & $\checkmark$& $\checkmark$ &  \extension{76.21} &  \extension{26.62}  &  \extension{77.19} & \extension{28.64} &  --  &   --  \\
  
 & & \textit{Ours (KL+Ent)} & $\checkmark$& $\checkmark$  &\extension{\underline{78.10}} & \extension{\textbf{23.64}}  &  \extension{\textbf{78.50}}  &  \extension{\underline{24.39}} &  \extension{78.45}  & \extension{23.34}  \\
\midrule
\multirow{5}{*}{\textit{Set-10}} & \multirow{4}{*}{Single Branch} & Lower bound  & $\checkmark$ & -- &  \extension{77.65} & \extension{22.45} &   --  &   -- \\ 
& & \textit{Single}  & $\checkmark$& $\checkmark$& \extension{79.14}  &
\extension{\textbf{18.22}} &  -- & -- &  -- & -- \\
 & &\rev{\textit{Single + Ent}} & $\checkmark$& $\checkmark$&  \rev{81.72} & \rev{22.99}  & -- &  -- &  -- & -- \\
 & &\rev{\textit{Upper Bound (Set-10)}} & $\checkmark$& --&  \rev{87.15} & \rev{11.08}  & -- &  -- &  -- & -- \\
  \cmidrule{2-11}
 & \multirow{3}{*}{Dual Branch} &
 \textit{Decoupled} \citep{luo2020semi}     & $\checkmark$ & $\checkmark$ &  \extension{79.18}  & \extension{20.16}  & \extension{18.11}  & \extension{61.53} &  -- & -- \\
 &  & \textit{Ours (KL)}  & $\checkmark$& $\checkmark$ &\extension{81.50} &  \extension{21.99} &    \extension{81.01}  &  \extension{23.05} &  -- & --  \\
 & & \textit{Ours (KL+Ent)}  & $\checkmark$& $\checkmark$ & \extension{\underline{81.79}} &  \extension{21.13} &   \bf \extension{83.05}  &  \extension{\underline{18.93}} &  \extension{82.36}  & \extension{20.08} \\
  \midrule
 All images & Single Branch & Upper bound  & $\checkmark$ & -- & \extension{91.30}  &  \extension{5.01}  &  -- & --  \\
  \bottomrule
\end{tabular}
}

\label{table:la-main}
\end{table*}

\paragraph{\extension{\textbf{Results on left-atrium (LA) segmentation}}}\extension{Beyond the ACDC dataset, we performed experiments on the more challenging LA segmentation task, whose results are reported in Table \ref{table:la-main}. These results align with the observations in the ACDC dataset (Table \ref{table:acdc-main}). In particular, the proposed approach outperforms consistently the different baselines across all the settings, with a significant gap in less supervised scenarios. For example, in \textit{Set-3}, our model brings a gain of 15\% in terms of DSC compared to the recent work in \citep{luo2020semi}, whereas the difference amounts to approximately 10 mm in terms of HD. On the other hand, there is a noticeable decline in this gap as the number of fully supervised samples increases. Nevertheless, the differences between our method and the best performing baseline are still remarkable, with nearly 4\% in terms of DSC. \rev{Furthermore, and similarly to the ACDC dataset, simply adding an entropy minimization term to the Single model does not translate into similar performances to those observed by the proposed model. This empirically \textit{i)} demonstrates that our method is not equivalent to a model with a single architecture integrating an entropy term and \textit{ii)} supports our hypothesis that decoupling the decoder network to avoid supervision interference yields better segmentation results.} 

Qualitative evaluation is visually assessed in Fig. \ref{fig:qualitative-la}, which depicts the segmentation results across models on the \textit{Set-3} and \textit{Set-5} settings. Similarly to the visual examples in Fig. \ref{fig:qualitative-main}, single models generate inconsistent segmentations, which result in both large under and over-segmentations. Even though decoupling single models in dual-stream architectures seem to reduce the amount of false positives, it typically comes at the price of failing to identify target regions. In contrast, both of our models provide a substantial improvement on the segmentation quality, with the model integrating the KL and the entropy terms providing the closest results to the ground-truth.}

\paragraph{\extension{\textbf{Model complexity}}}

\extension{Last, we evaluate the model complexity, measured in number of parameters for the different analyzed methods (Table \ref{table:complex}). Several methods \citep{Peng2021boost} employ a single model, whereas other approaches \citep{yu2019uncertainty} need to duplicate this into a teacher-student architecture. \rev{It is important to realize that some of these models, e.g., \citep{yu2019uncertainty}, have only half of reported parameters to be learned by gradient descent, since the teacher parameters are updated via exponential average moving. Nevertheless, as the parameter values need to be also stored, we have included them in our calculation.} In terms of complexity, our model lies in between these two strategies, as despite integrating two decoupled decoders, the encoder is shared among the two branches. On the other hand, the closest method in terms of performance, i.e., \citep{wang2019mixed}, comes at the price of significant complexity increase, which may hinder its deployment in realistic scenarios. }

\begin{table}[h!]
\scriptsize
\centering
  \caption{\extension{Model complexity in terms of parameters during training \rev{and inference time.}}}
\resizebox{1.0\columnwidth}{!}
{
\begin{tabular}{lcc}
\toprule

    Model    &  $\#$Params & \rev{ Time (per sample) (ms)}\\
\midrule
\extension{Single Branch} & \extension{31,042,434} & \rev{4.9} \\
\extension{GLMI ~\citep{Peng2021boost}} & \extension{31,042,434} & \rev{4.9}\\
\extension{Dual Branch ~(Ours)} & \extension{41,137,220} & \rev{7.1}\\
\extension{UA-MT ~\citep{yu2019uncertainty}} & \extension{62,084,868} & \rev{9.9} \\
\extension{SSCO ~\citep{Wang2020selfpaced}} & \extension{248,339,472} & \rev{37.6} \\

  \bottomrule
\end{tabular}
}
\label{table:complex}
\end{table}

\section{\extension{Conclusion}}

\extension{In this work we have presented a novel formulation for semantic segmentation under the mixed-supervised paradigm. In addition to the standard cross-entropy loss over the labeled pixels, we integrate two important terms in the global learning objective, which have demonstrated to bring a significant boost in performance. First, a Shannon entropy loss defined over the less-supervised images encourages confident predictions on these images. Secondly, a KL divergence transfers the knowledge from the predictions generated by the strongly supervised branch to the less-supervised branch. As shown in the experiments, the latter term plays a crucial role in our global learning objective, as it serves as a strong prior for the bottom branch, avoiding trivial solutions resulting from the entropy term.}

\extension{Furthermore, we have discussed an interesting link between Shannon-entropy minimization and standard pseudo-mask generation, typically used in semi and weakly supervised semantic segmentation. Motivated from a gradient dynamics perspective, we further argue that the former should be preferred over the latter to leverage information from unlabeled pixels. In particular, we show that plugging pseudo-masks in a cross-entropy loss is equivalent to minimizing the min-entropy (Fig \ref{fig:entropy}). Hence, for uncertain predictions, i.e., in the middle of the simplex, the gradient is much higher than with the entropy, pushing the predictions at the beginning of the training towards the vertex. In addition, we provide empirical evidence that employing pseudo-labels has this undesired effect.}

\extension{Through extensive experiments, we have rigorously assessed the impact of the different elements of the proposed formulation. Our experiments have further confirmed the usefulness of our method on two publicly available segmentation benchmarks. We have also demonstrated the significant superiority of our approach compared, not only to existing literature in mixed-supervised segmentation, but also to well-known recent semi-supervised methods. It is worth mentioning that, even though our method requires slightly more supervision, the cost of obtaining it is negligible. Furthermore, compared to similar performing methods that require training multiple models, our approach is substantially less complex in terms of number of parameters.}

\extension{The proposed framework is straightforward to use, does not incur in significant computational costs and can be used with any segmentation network architecture or segmentation loss. Future work will address the integration of other types of supervision in the bottom branch, for example in the form of image tags or anatomical priors. }

\section*{Acknowledgements}
This work is supported by the National Science and Engineering Research Council of Canada (NSERC), via its Discovery Grant program. We also thank Calcul Quebec and Compute Canada.




        %

\bibliographystyle{model2-names.bst}\biboptions{authoryear}
\bibliography{main}

\newpage
      
\appendix


\setcounter{section}{0}
\setcounter{table}{0}
\setcounter{figure}{0}

\section{\rev{Parameters search}}
\label{sec:paramsearch}

\rev{The proposed model contains several objective terms, each balanced with a different weighting factor. Thus, we sequentially found the best value for each balancing term, and with this term fixed we moved to the next term. We report below the intermediate steps to find the best set of hyperparameters, and their corresponding results.}

\subsection{\rev{Sensitivity of $\lambda_{w}$ on the \textit{Single} model.}}

\rev{To find the best set of hyperparemeters, we first searched the optimal value of $\lambda_{w}$ in the \textit{Single} model, which was then fixed to find the remaining hyperparameters. The results from the conducted experiments are reported in Table \ref{tab:results-acdc-ablationW_sup}. We can observe that, similarly to the model trained with the whole learning objective, setting $\lambda_{w}=0.001$ provided the best results consistently across the three settings.}

\begin{table}[h!]
\scriptsize
\centering
\caption{\rev{Impact of $\lambda_{w}$ on the \textit{Single} model. \revB{Results on the ACDC dataset}.}}
\resizebox{0.9\columnwidth}{!}
{
\begin{tabular}{l|cc|cc|cc|}
\toprule
         & \multicolumn{2}{c}{\textit{Set-3}} & \multicolumn{2}{c}{\textit{Set-5}} & \multicolumn{2}{c}{\textit{Set-10}} \\
         \midrule
         & DSC        & HD-95       & DSC        & HD-95       & DSC        & HD-95        \\
\midrule
$\lambda_{w}=1$  & \rev{22.93} &  \rev{125.61} & \rev{64.21} & \rev{64.31} &   \rev{69.55}&   \rev{53.10}\\
$\lambda_{w}=0.1$  & \rev{45.20} & \rev{92.37}  & \rev{63.74} & \rev{52.01} &  \rev{75.35}  &   \rev{50.52}\\
$\lambda_{w}=0.01$  & \rev{55.04} &  \rev{78.15} & \rev{65.38} & \rev{52.59} & \rev{76.94} &  \rev{48.53} \\
$\lambda_{w}=0.001$   & \bf \rev{57.42} & \bf \rev{78.80}  & \bf \rev{70.73} &  \bf \rev{51.34} & \bf \rev{78.17} & \bf \rev{42.99} \\
$\lambda_{w}=0.0001$   & \rev{41.82} &  \rev{91.17} & \rev{60.93} & \rev{51.93} & \rev{74.18} & \rev{50.98} \\

\bottomrule
\end{tabular}
}
\label{tab:results-acdc-ablationW_sup}
\end{table}

\subsection{\rev{Sensitivity of $\lambda_{KL}$ on the \textit{KL} model.}}

\rev{Once the optimal value for $\lambda_{w} (\lambda_{w}=0.001)$ is found, we optimize the $\lambda_{KL}$ hyperparameter. In particular, with $\lambda_{w}$ fixed, we evaluate the performance of the $KL$ model across different values of $\lambda_{KL}$. The optimal value found will be used in our final model, which also includes the entropy term into the learning objective. We can observe that both $\lambda_{KL}=50$ and $\lambda_{KL}=100$ achieve similar results in terms of DSC. Nevertheless, when the value of $\lambda_{KL}$ is fixed  to 50, the values of the HD metric decrease, suggesting that it is a better value for the $KL$ term. Hence, we will fix this hyperparameter to 50 in the whole proposed model.}

\begin{table}[h!]
\scriptsize
\centering
\caption{\rev{Impact of $\lambda_{KL}$ on the \textit{KL}-based model. \revB{Results on the ACDC dataset}.}}
\resizebox{0.9\columnwidth}{!}
{
\begin{tabular}{l|cc|cc|cc|}
\toprule
         & \multicolumn{2}{c}{\textit{Set-3}} & \multicolumn{2}{c}{\textit{Set-5}} & \multicolumn{2}{c}{\textit{Set-10}} \\
         \midrule
         & DSC        & HD-95       & DSC        & HD-95       & DSC        & HD-95        \\
\midrule
$\lambda_{KL}=0.1$  & \rev{63.37} &  \rev{87.63} & \rev{74.26} & \rev{44.63} &   \rev{84.12}&   \rev{32.09}\\
$\lambda_{KL}=1$  & \rev{65.70} & \rev{79.77}  & \rev{75.21} & \rev{52.29} &  \rev{87.74}  &   \rev{21.10}\\
$\lambda_{KL}=10$  & \rev{64.12} &  \rev{66.47} & \rev{75.56} & \rev{37.45} & \rev{87.04} &  \rev{27.61} \\
$\lambda_{KL}=50$   & \bf \rev{71.41} & \bf \rev{61.37}  & \bf \rev{81.79} &  \bf \rev{24.07} & \bf \rev{89.20} & \bf \rev{9.78} \\
$\lambda_{KL}=100$   & \rev{71.23} &  \rev{61.92} & \rev{81.52} & \rev{27.71} & \rev{88.66} & \rev{12.01} \\
$\lambda_{KL}=1000$   & \rev{63.64} &  \rev{77.40} & \rev{78.60} & \rev{26.71} & \rev{86.65} & \rev{14.05} \\
\bottomrule
\end{tabular}
}
\label{tab:results-acdc-ablationKL_sup}
\end{table}

\subsection{\revA{Single vs Double Branch models}}

\revA{We now perform further experiments on the impact of the entropy weighting factor on the single model. These results, which are reported in Table \ref{tab:results-entropy}, show that the value selected for the entropy term is actually a good compromise across settings. As this value increases, the performance is degraded due to the entropy term dominating the learning and driving the results towards trivial solutions. On the other hand, if the balancing weight is small, it resembles to the baseline approach. Note that, however, the results obtained by this model (Single + Entropy) are not comparable to the performance of the proposed formulation, as suggested by the reviewer.}

\begin{table}[h!]
\scriptsize
\centering
\caption{\revA{Impact of $\lambda_{Ent}$ on the Single-based model that integrates only the entropy as additional term, whose overall learning objective is $\mathcal{L}_s + \lambda_w \mathcal{L}_w + \lambda_{ent} \mathcal{H} (\mathbf{p})$. \revB{Results on the ACDC dataset}.}}
\begin{tabular}{l|cc|cc|}
\toprule
         & \multicolumn{2}{c}{\revA{\textit{Set-3}}} & \multicolumn{2}{c}{\revA{\textit{Set-5}}}  \\
         \midrule
         & \revA{DSC}        & \revA{HD-95}       & \revA{DSC}        & \revA{HD-95}           \\
\midrule
\revA{$\lambda_{Ent}=0.0$}  & \revA{57.42} &  \revA{78.80} & \revA{70.73} & \revA{51.34} \\
\revA{$\lambda_{Ent}=0.1$}  & \revA{51.32} &  \revA{84.45} & \revA{69.62} & \revA{48.53} \\
\revA{$\lambda_{Ent}=1$}  & \revA{43.01} & \revA{83.98}  & \revA{74.92} & \revA{55.24} \\
\revA{$\lambda_{Ent}=10$}  & \revA{38.61} &  \revA{100.86} & \revA{62.66} & \revA{54.34}  \\

\bottomrule
\end{tabular}
\label{tab:results-entropy}
\end{table}

\section{\rev{Parameter search for semi-supervised methods}}
\label{sec:paramsearch-semi}

\rev{During the validation of our model we found that the default hyperparameters of compared methods were not optimal in our setting. Thus, for a fair comparison, we searched the optimal hyperparameters on ACDC dataset for each semi-supervised method as well, and reported the best scores (shown in Table \ref{table:ssl} of the main text). Here, we include the intermediate steps to find the best set of hyperparameters for each method and their corresponding results. Note that we tune the parameters under the Set-3 setting and use the best findings for the other two settings, i.e. Set-5 and Set-10.}

\subsection{\rev{UA-MT}}

\rev{The formulation of UA-MT \citep{yu2019uncertainty} consists of two terms, i.e., the supervised loss and the unsupervised consistency loss for measuring the consistency between the prediction of the teacher and the student model, with a balancing consistency weight $\lambda_c$ controlling their relative contribution.
We searched the optimal $\lambda_c$ and reported the corresponding results in Table \ref{tab:results-acdc-uamt}, where we can see that $\lambda_{c}=0.1$ provided the best results.} 

\begin{table}[h!]
\scriptsize
\centering
\caption{\rev{Impact of the consistency weight ($\lambda_c$) on UA-MT \citep{yu2019uncertainty}.
The study is conducted on the ACDC dataset (Set-3 setting).
}}
\label{tab:results-acdc-uamt}
\begin{tabular}{l|cc}
\toprule
         & DSC  & HD-95 \\
\midrule
\rev{$\lambda_{c}=0.01$} & \rev{67.90} & \rev{42.57}   \\
\rev{$\lambda_{c}=0.1$}  & \bf \rev{70.62} & \bf \rev{38.06} \\
\rev{$\lambda_{c}=1$}  & \rev{68.14} & \rev{45.01}  \\
\rev{$\lambda_{c}=5$}  & \rev{67.31}  & \rev{54.96}  \\
\rev{$\lambda_{c}=10$}    & \rev{67.45}  & \rev{43.06}  \\
\bottomrule
\end{tabular}
\end{table}

\subsection{\rev{GLMI}}

\rev{The formulation of GLMI \citep{Peng2021boost} includes a composite loss with four terms, i.e., supervised loss, the mutual information loss on global feature embedding and local feature embedding, and the consistency loss on different transformation of the same given input. Hence, three hyper-parameters are introduced for tuning the relative contributions of the last three terms, i.e., $\lambda_{MI}^{global}$, $\lambda_{MI}^{local}$ and $\lambda_{cons}$.
In our implementations, the best hyper-parameter setting was slightly different from that suggested in \citep{Peng2021boost}. The reason for this could be that we focus on the task of left ventricular (LV) endocardium segmentation and the potential difference on the experimental environment. Specifically, we empirically fixed $\lambda_{cons}$ to $10$ and report the results of different $\lambda_{MI}^{global}$ and $\lambda_{MI}^{local}$, as shown in Table \ref{tab:results-acdc-glmi}. The best result is achieved by setting $\lambda_{MI}^{global}$ and $\lambda_{MI}^{local}$ to $1$ and $0.1$ respectively.
}

\begin{table}[h!]
\scriptsize
\centering
\caption{\rev{Impact of $\lambda_{MI}^{global}$ and $\lambda_{MI}^{local}$ on GLMI \citep{Peng2021boost}.
Dice score is reported for each setting and the study is conducted on ACDC dataset (Set-3).
}}
\label{tab:results-acdc-glmi}
\begin{tabular}{l|ccc}
\toprule
       & \rev{$\lambda_{MI}^{local}=0.05$}  & \rev{$\lambda_{MI}^{local}=0.1$}  & \rev{$\lambda_{MI}^{local}=1$}  \\
\midrule
\rev{$\lambda_{MI}^{global}=0.1$}  & \rev{70.61}  & \rev{70.07} & \rev{69.99}  \\
\rev{$\lambda_{MI}^{global}=1$}  & \rev{74.31}  & \rev{\bf 76.27} & \rev{72.85} \\
\revA{$\lambda_{MI}^{global}=2$}  & \revA{73.76}  & \revA{75.79} & \revA{74.70} \\
\bottomrule
\end{tabular}
\end{table}

\subsection{\rev{SSCO}}

\rev{In SSCO \citep{Wang2020selfpaced}, two hyper-parameters, i.e., $\lambda_1$ and $\lambda_2$, are introduced to balance the relative contributions of the self-paced co-training loss and the self-consistency loss.
We tune the two hyper-parameters based on the values reported in the original paper \citep{Wang2020selfpaced} and found the values reported in the paper ($\lambda_1 = 0.5, \lambda_2=4$) provided best scores in our implementation as well.
Detailed comparison results are shown in Table \ref{tab:results-acdc-ssco}
}

\begin{table}[htb!]
\scriptsize
\centering
\caption{\rev{Impact of $\lambda_1$ and $\lambda_2$ on SSCO \citep{Wang2020selfpaced}.
Dice score is reported for each setting, and the study is conducted on ACDC dataset (Set-3).
}}
\label{tab:results-acdc-ssco}
\begin{tabular}{l|ccc}
\toprule
         & \rev{$\lambda_2=1$}  & \rev{$\lambda_2=4$} & \rev{$\lambda_2=8$} \\
\midrule
\rev{$\lambda_1=0.1$}  & \rev{75.08}  & \rev{74.80} & \rev{73.38}  \\
\rev{$\lambda_1=0.5$}  & \rev{74.71} & \bf \rev{77.16} & \rev{75.95} \\
\revA{$\lambda_1=1 $}  & \revA{74.01} & \revA{75.68} & \revA{74.83} \\
\bottomrule
\end{tabular}
\end{table}

\section{\rev{Additional failure cases from Pseudo-Labels}}

\begin{figure}[h!]
    \centering
    \includegraphics[width=0.475\textwidth]{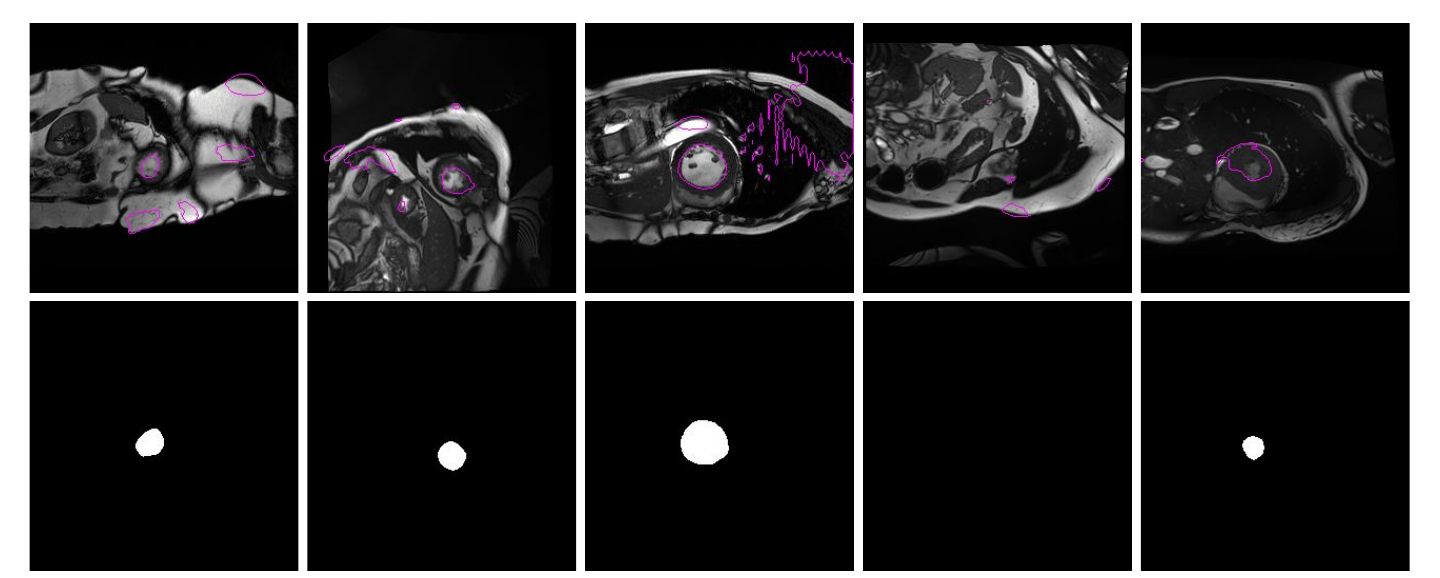}
    \caption[]{\rev{Additional failure cases (\textit{Set-3}) which are employed as pseudo-labels in the \textit{proposals-based} approach \rev{(top)}, whose errors are reinforced during training, \rev{and their corresponding ground truth}. Best viewed in colours.}}
    \label{fig:failure_Proposals-sup}
\end{figure}

\rev{In this section we show additional failure cases obtained from the \textit{Pseudo-labels} approach. As shown in Figure~\ref{fig:failure_Proposals-sup}, It is important to mention that these pseudo-labels are used in subsequent iterations to train the deep network, which brings a high risk of propagating these errors through the training. These visual results, which are supported by the quantitative evaluation in the main paper, support our hypothesis that minimizing entropy should be preferred over the use of pseudo-labels, as the errors are propagated through the training.}
 
\section{\rev{Qualitative comparison with the upperbound}}

\begin{figure*}[h!]
    \centering
    \includegraphics[width=1\textwidth]{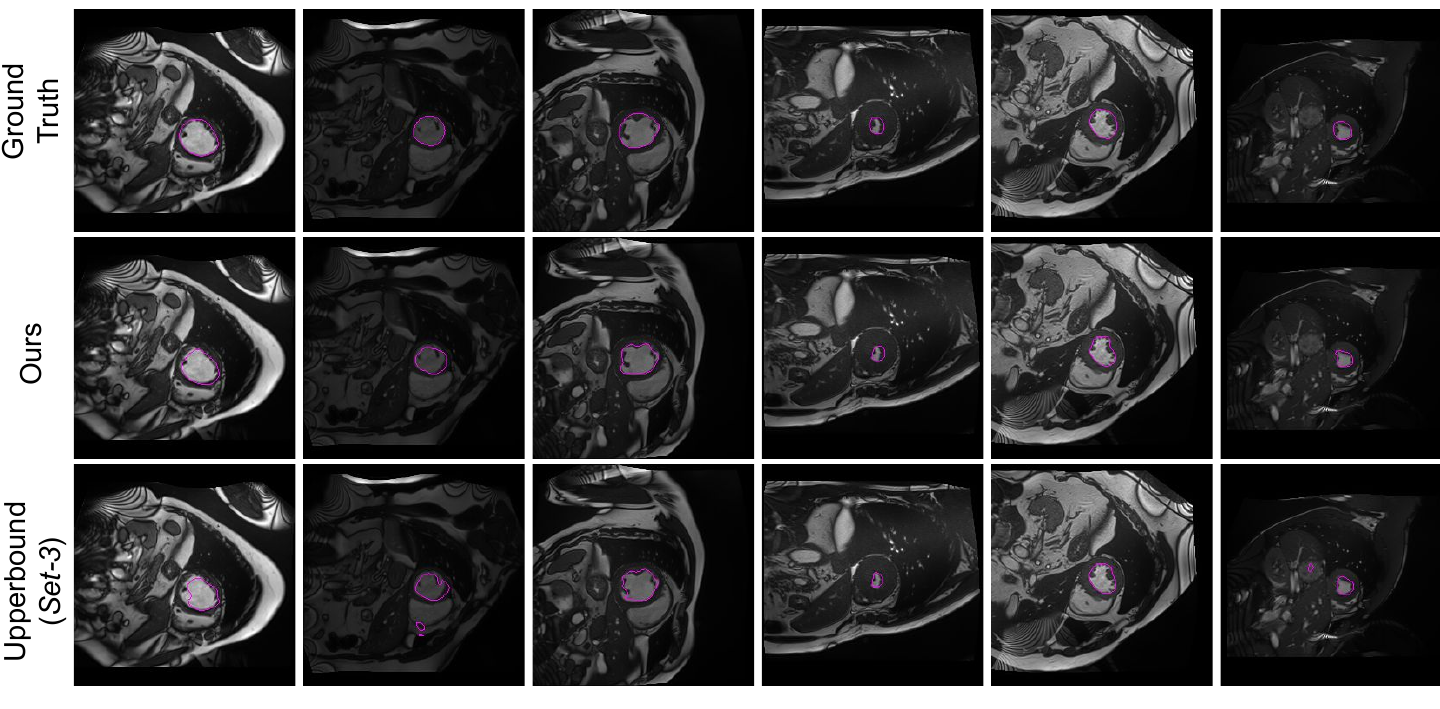}
    \caption[]{\rev{Qualitative results on the Test dataset of \textit{Set-3} obtained by our model (\textit{middle}) and the upperbound model (\textit{top}) on the same dataset. The corresponding ground truth is depicted in the (\textit{top}) row. }}
    \label{fig:qualitative-set3}
\end{figure*}

\begin{figure*}[h!]
    \centering
    \includegraphics[width=1\textwidth]{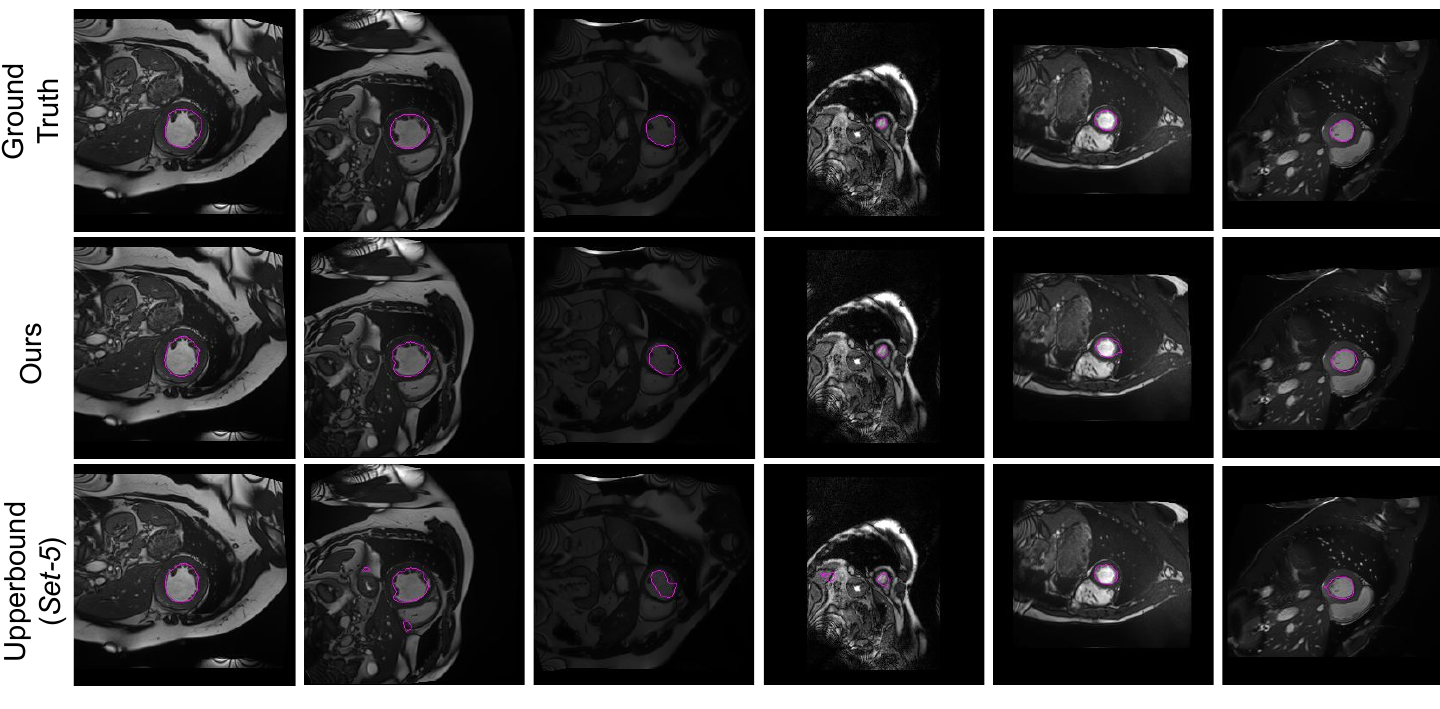}
    \caption[]{\rev{Qualitative results on the Test dataset of \textit{Set-5} obtained by our model (\textit{middle}) and the upperbound model (\textit{top}) on the same dataset. The corresponding ground truth is depicted in the (\textit{top}) row. }}
    \label{fig:qualitative-set5}
\end{figure*}

\rev{We depict in Figures \ref{fig:qualitative-set3} and \ref{fig:qualitative-set5} several visual results on the \textit{Set-3} and \textit{Set-5} from our method compared to the individual upperbounds. In this scenario, both models are trained with the same images, being the only difference the level of supervision provided. We can observe that despite the slight quantitative differences in Table \ref{table:acdc-main} of the main paper, visual results indicate that the proposed method can achieve very similar results to the upperbound model, sometimes generating more similar segmentations to the ground truth.} 

\end{document}